\documentclass[onecolumn]{aastex6}
\usepackage{setspace}
\collaborationName{Friends of AASTeX}

\begin{document}


\title{Collisional growth of icy dust aggregates in disk formation stage: difficulties for planetesimal formation via direct collisional growth outside the snowline}


\author{Kenji Homma\altaffilmark{1} and Taishi Nakamoto\altaffilmark{1}}
\affil{$^1$Department of Earth and Planetary Sciences, Tokyo Institute of Technology, Meguro, Tokyo, 152-8551, Japan; nakamoto@eps.sci.titech.ac.jp}
\and



\begin{abstract}
Highly porous dust aggregates can break through the radial drift barrier, but previous studies assumed disks in their later stage, where the disks have a very small mass and low temperature. 
 In contrast, dust coagulation should begin in the very early stage such as the disk formation stage because the growth timescale of dust is shorter than the disk formation timescale if there is no process to suppress the collisional growth of dust. 
We investigate the possibility of planetesimal formation via direct collisional growth in the very early stage of a protoplanetary disk. 
We show that, in the very early stage of protoplanetary disks, icy dust aggregates suffer radial drift and deplete without forming planetesimal-sized objects.  
 This is because as the disk temperature easily increases by the viscous heating in the disk formation stage, the area where the dust can break through the radial drift barrier is restricted only to the inside the snowline.
This suggests that in the disk formation stage,
icy planetesimal formation via direct collisional growth of dust is difficult.
\end{abstract}




%
%

\section{Introduction}

Planets are thought to be formed through collisions of objects called ``planetesimals'' in a protoplanetary disk.
Planetesimals are formed from dust particles in the disk, though their formation processes are still under debate.
Some proposed planetesimal formation mechanisms include the gravitational instability of the dust layer \citep{goldreich1973},
streaming instability \citep[e.g., ][]{johansen2007b},
and the successive growth of dust particles by mutual collisions \citep{okuzumi2012, arakawa2016}.

These planetesimal formation processes are confronted with some problems. 
Turbulence in the disk stirs the dust and prevents dust from sufficiently settling to cause a gravitational instability \citep[e.g., ][]{cuzzi1993}.  
A streaming instability needs an enhancement in the dust-to-gas mass ratio \citep{johansen2009}. 
In planetesimal formation via direct collisional growth, the radial inward drift of macroscopic dust aggregates due to the gas drag in protoplanetary disks \citep{adachi1976,weidenschilling1977} is a serious problem called  the ``radial drift barrier." 
In any mechanism, the collisional growth of dust particles may play an important role.

Recent theoretical studies show that the evolution of the internal structure of dust aggregates is key for dust aggregate growth. 
For example, $N$-body simulations \cite[e.g.,][]{wada2008,suyama2008,suyama2012} reveal
the evolution of the internal structure and the strength of  aggregates for collisional compression. 
The strength of highly porous aggregates for static compression was investigated by \cite{kataoka2013a}. 
These studies show that aggregates do not have a compact structure but become a fluffy structure with their growth. 
 
These studies have helped in forming planetesimals via direct collisional growth.
\cite{okuzumi2012} investigated the collisional growth of icy dust aggregates and their porosity evolution using the recipe of \cite{suyama2012} including collisional compression. 
They showed that aggregates have a very high porosity by growing with a similarly sized collision, which is called ballistic cluster--cluster aggregation (BCCA). 
They also showed that highly porous icy dust aggregates can grow to planetesimals without radial drift inside $10 \ {\rm AU}$ via direct collisional growth because the highly porous aggregates have a higher collision rate than compact aggregates. 
\cite{kataoka2013b} showed that icy aggregates can grow to planetesimals by collisional growth even if aggregates suffer static compression (i.e., gas compression in protoplanetary disks and the self-gravity of aggregates).
 
These studies \citep{okuzumi2012, kataoka2013b}, however, assumed disks in their later stage, where the disks have a very small mass and low temperature, and set the initial condition that all icy dust particles are present as micron-sized dust  particles with a dust-to-gas mass ratio of 0.01 in the entire disk. 
In contrast, dust coagulation should begin at the same time as disk formation if there is no process to suppress the collisional growth of dust. 
If dust grows and radial drift occurs in the disk formation stage, the dust will be depleted, and it will be difficult to form planetesimals after that.
It seems that the possibility of planetesimal formation via direct collisional growth at a very early phase of a protoplanetary disk should be examined, and that is the purpose of this study.
 
 In order to simplify the problem, in the same way as \cite{okuzumi2012}, we ignore some processes that suppress the collisional growth of aggregates including bouncing \citep[e.g., ][]{zsom2010, zsom2011}, collisional fragmentation \citep[e.g., ][]{birnstiel2009, birnstiel2010, birnstiel2012}, erosion \citep[e.g.,][]{krijt2015}, and the Coulomb interaction \citep[e.g.,][]{okuzumi2009a, okuzumi2011a, okuzumi2011b}. 
Although bouncing is often observed in laboratory experiments, $N$-body numerical experiments show that bouncing is unlikely to occur when aggregates have porosity \citep{wada2011}. 
Collisional fragmentation is serious when we consider rocky dust particles, but it is considerably relieved in the case of icy dust. 
In the case of rocky dust, the impact velocity experiencing catastrophic fragmentation is estimated to be a few ${\rm m \ s^{-1}}$ from both laboratory and $N$-body experiments \citep{blum2008, wada2009}. 
However, for icy dust, it is up to $35-70 \ {\rm m \ s^{-1}}$ estimated from $N$-body simulations \citep{wada2009} assuming a $0.1$-${\rm \mu m}$-sized icy monomer. 
For this reason, in this study, we focus on the collisional growth of icy dust outside the snowline instead of ignoring collisional fragmentation.
{
Dust aggregates can also lose mass through erosion which is caused by high-velocity collisions with small dust particles/aggregates \citep[e.g., ][]{schrapler2011, seizinger2013}.
The critical velocity for the erosion, however, is suggested to be 100 m s$^{-1}$ or higher if monomers are 0.1 $\mu$m-sized icy particles \citep{gundlach2015}.
Thus, we ignore this process as well as catastrophic fragmentation for simplicity. We will discuss the validity of this assumption in Section \ref{sec:possibility}.
}

The Coulomb force cannot be ignored for negatively charged dust, and the Coulomb interaction can slow the initial dust growth, which is called the ``charge barrier" \citep{okuzumi2009a}. 
However, this process is very complicated and is not clearly understood yet; thus, for the sake of simplicity, we ignore the Coulomb interaction in this study, but we discuss the importance of this process in Section \ref{sec:possibility}.

The formation of protoplanetary disks has been studied by hydrodynamical simulations \citep[e.g., ][]{yorke1993, machida2010} and cylindrical 1-D simulations of the disk evolution \citep[e.g., ][]{nakamoto1994, hueso2005}. 
Such disks have a mass supply of gas and dust from the envelope, and their lifetime is considered to be approximately 0.5 Myr.
An important feature in such a stage is that the disk becomes heavier to show the gravitational instability due to the mass supply from the envelope \citep[e.g., ][]{nakamoto1994, tsukamoto2015}.
In addition, the high disk mass accretion rate to the central star makes viscous heating more effective, and the temperature of the disk becomes sufficiently high such that the snowline reaches $10 {\rm AU}$ \citep{zhang2015}.
The increases in the mass and temperature of the disk may affect the behavior of gas drag to dust and the radial drift speed of dust.
Therefore, the disk in the formation stage is greatly different from the disk that \cite{okuzumi2012} assumed.
 
There are some studies that investigated the collisional growth of dust in the disk formation stage.
\cite{birnstiel2010} investigated the gas and dust evolution including the mass accretion from the molecular cloud core and showed that no planetesimal forms in the disk formation stage. 
However, they assumed that dust has a compact structure and did not consider the internal density evolution of aggregates, although dust aggregates with a high porosity increase the collisional growth rate. 
In a complementary work, \cite{tsukamoto2017} investigated the highly porous dust growth in gravitationally unstable disks with mass accretion from the envelope. 
However, they did not consider the gas drag law for dust aggregates with a high Reynolds number, although macroscopic dust has a large Reynolds number in the disks that they used as the model. 
They may have overestimated the growth rate of dust since the growth rate with dust at a high Reynolds number gives the maximum value of the growth rate \citep{okuzumi2012}. 
In addition, they did not solve the evolution of the dust size distribution; thus, the internal density of dust was treated as a model parameter. 
The supply of dust from the envelope affects the dust size distribution, and the internal density evolution of the aggregate may be different from BCCA. 

In this study, we simulate the evolution of the radial size distribution of dust and the gas surface density in a disk simultaneously considering the mass accretion from the molecular cloud core. 
Unlike previous studies \citep{birnstiel2010, tsukamoto2017}, we also calculate the internal density evolution of aggregates from the size of the colliding aggregates. 
We use the method used in \cite{nakamoto1994} and \cite{hueso2005} as the gas disk evolution. 
In addition, we use the method developed by \cite{okuzumi2009, okuzumi2012}, which allows for the calculation of the radial size evolution of dust and the evolution of the average volume of aggregates at each orbital radius and size. 

This paper is organized as follows. 
In Section \ref{sec:model}, the models of the gas and dust disk evolution are described. 
Our calculation results are presented in Section \ref{sec:result}. 
A semianalytical understanding of the results, the validity of our model, and future prospects are discussed in Section \ref{sec:discussion},
and a summary of this study is presented in Section \ref{sec:sum}.

%
%
\section{MODEL}
\label{sec:model}
In this study,  the coagulation of icy dust aggregates and their radial transport in a protoplanetary disk is investigated, taking into account the mass accretion to the disk from the collapsing molecular cloud core.
First, the gas disk evolution model including the infall from the molecular cloud core is introduced in Section \ref{sec:gas}.
Then, the dust evolution model including the collisional growth, the global transport in the protoplanetary disk, and the internal density evolution of dust aggregates is described in Section \ref{sec:dust}.

Dust particles generally influence the dynamics of the gas in the disk through the dust--gas interaction due to the gas drag force, especially when the dust spatial mass density is close to or larger than that of the gas.
Moreover, the size evolution of dust aggregates affects the opacity of the disk, and the opacity may influence the disk temperature. 
In our model, however, these effects on the gas disk are ignored for simplicity.

Cylindrical coordinates $(r, \phi, z)$ are used to describe the phenomena in a disk.
The central star is located at the origin, and the disk midplane is in the $z=0$ plane.
It is also assumed that the system is axially symmetric.

%
%
\subsection{Disk Model}
\label{sec:gas}
Our model of the gas disk evolution follows the models described by \cite{nakamoto1994} and \cite{hueso2005}.

%
%
\subsubsection{Molecular Cloud Core Collapse}
The evolution of a disk depends on the initial infall phase associated with the molecular cloud core collapse. 
This phase is still not clearly understood; therefore, in order to simplify the problem, 
the infall model by \cite{shu1977}, in which the molecular cloud core is assumed to be isothermal and spherically symmetric, is adopted in this study.
It was shown that the molecular cloud core undergoes inside-out collapse and the mass accretion rate from the molecular cloud core $\dot{M}$ is given by$
\dot{M}=0.975 \frac{c_{\rm s,cd}^3}{G}
$,
where $G$ is the gravitational constant,
$c_{\rm s,cd}=(k_{\rm B}T_{\rm cd}/m_{\rm g})^{1/2}$ is the isothermal sound speed in the molecular cloud core, $T_{\rm cd}$ is the temperature of the molecular cloud core,
$k_{\rm B}$ is the Boltzmann constant,
and $m_{\rm g}$ is the mean mass of a gas molecule \citep{shu1977}.
The temperature of the cloud core, $T_{\rm cd}$, is typically $10-20$ K \citep{van1993},
and it is regarded as a model parameter in this study.

The infall materials in a spherical shell in the molecular cloud core fall inside the centrifugal radius $r_{\rm c}$. 
The place on the disk where the infalling material lands depends on the specific angular momentum of the infalling material. 
Assuming that the molecular cloud core initially rotates as a rigid body and assuming the conservation of the angular momentum, the balance between gravity and the centrifugal force leads to the centrifugal radius at $t$, $r_{\rm c}(t)$, as
$r_{\rm c}(t)={l(t)^4 \omega_{\rm cd}^2}/\{{GM(t)}\}$, 
where $l(t)$ is the distance from the origin to the initial position in the molecular cloud core of the material, which reaches the disk at $t$;
$\omega_{\rm cd}$ is the initial angular velocity of the molecular cloud core; and $M(t)$ is the total mass of star--disk system at the time $t$. 
The angular velocities of molecular cloud cores are estimated from the observations of velocity gradients in clouds \citep{goodman1993}, and their typical values range from  $10^{-15} \  {\rm s^{-1}}$ to $10^{-13} \  {\rm s^{-1}}$. 
In this study,  $\omega_{\rm cd}$ is regarded as a model parameter.

The collapse solution yields $l(t)=c_{\rm s,cd}t/2$, and \cite{hueso2005} wrote the centrifugal radius $r_{\rm c}$ as
\begin{eqnarray}
\label{eq:eq6}
r_{\rm c}(t)=53\biggl(\frac{\omega_{\rm cd}}{10^{-14}{\rm s^{-1}}}\biggr)^2\biggl( \frac{T_{\rm cd}}{10 {\rm K}}\biggr)^{-4} \biggl( \frac{M(t)}{1M_{\odot}} \biggr)^3 \  {\rm AU}.
\end{eqnarray}
Assuming that the infalled materials are adopted by the disk at the orbital radius where their specific angular momentum corresponds to the angular momentum of the circular Kepler motion (this picture slightly differs from those of \cite{cassen1981} and \cite{nakamoto1994}), the mass accretion rate from the molecular cloud core to the unit surface area of the disk,
$S_{\rm g}(r,t)$, is given by 
\begin{eqnarray}
\label{eq:eq7}
  S_{\rm g}(r,t)=
  \left\{
  \begin{array}{cc}
     \frac{\dot{M}}{8 \pi r_{\rm c}^2} \biggl(\frac{r}{r_{\rm c}} \biggr)^{-3/2} \biggl[1- \biggl( \frac{r}{r_{\rm c}} \biggr)^{1/2} \biggr]^{-1/2} & (r<r_{\rm c}) \\
    0 & (r>r_{\rm c}) .
  \end{array}
  \right.
\end{eqnarray}

%
%
\subsubsection{Viscous Evolution of the Gas Disk}

Protoplanetary disks are geometrically thin; therefore, the temporal and spatial evolutions of the disk surface density
are examined, and the structure in the disk along the $z$ direction is not solved directly but
is assumed to be in the equilibrium state.
The time evolution of the gas surface density is described by the equation of continuity: 
\begin{eqnarray}
\label{eq:eq1}
\frac{\partial \Sigma_{\rm g}(r)}{\partial t}=-\frac{1}{r}\frac{\partial}{\partial r}(r v_{\rm g, r}(r) \Sigma_{\rm g}(r) )+S_{\rm g}(r),
\end{eqnarray}
where $\Sigma_{\rm g}$ is the gas surface density at the radius $r$, and $v_{\rm g}$ is the radial velocity of the gas.
The second term on the right-hand side, $S_{\rm g}$, 
is the source term that includes the infall materials from the molecular cloud core.
The radial velocity $v_{\rm g}$ is given by \citep{lynden-bell&pringle1974} 
\begin{eqnarray}
\label{eq:eq2}
v_{\rm g, r}=-\frac{3}{\Sigma_{\rm g}\sqrt{r}} \frac{\partial}{\partial r}(\Sigma_{\rm g} \nu_{\rm g} \sqrt{r}),
\end{eqnarray}
and $\nu_{\rm g}$ 
is the gas viscosity. 
The gas viscosity is assumed to be caused by the turbulence in the disk to explain the mass accretion to the central star. 
In this case, using the non dimentional parameter $\alpha$,
the gas viscosity is described as $\nu_{\rm g}=\alpha c_{\rm s}^2 \Omega$ \citep{shakura1973},
where $\Omega$ is the Kepler angular velocity, and $c_{\rm s}$ is the isothermal sound velocity
given by  $c_{\rm s}=(k_{\rm B}T/m_{\rm g})^{1/2}$, where $T$ is the gas temperature of the disk. 
The mean molecular mass is $m_{\rm g}=3.9 \times 10^{-24} \ {\rm g}$ when the mixing of ${\rm H}_2$ and He gases
is taken into consideration. 
Although the value of $\alpha$ is not clear, the accretion rates of T Tauri stars are compatible with $\alpha \simeq 10^{-2} $ \citep{hartmann1998}.

If the disk is gravitationally unstable, large-scale angular momentum transport due to the formation of spiral arms may occur. 
The stability of the disk is measured by Toomre's ${\cal Q}$ value defined by
${\cal Q}={c_{\rm s}\Omega}/ \{{\pi G \Sigma_{\rm g}}\}$ \citep{toomre1964}.
Gravitationally stable disks have a larger ${\cal Q}$,
and disks become marginally unstable when ${\cal Q} \simeq 2$. 
To take into account this angular momentum transport by the gravitational instability,
using the recipe of \cite{armitage2001}
the parameter $\alpha$ is modified as
\begin{eqnarray}
\label{eq:alpha}
\alpha(r)=\alpha_{\rm turb} + 0.01\Biggl( \biggl(\frac{{\cal Q}_{\rm cr}}{{\rm min}({\cal Q}_{\rm cr},{\cal Q}(r))}\biggr)^2-1 \Biggr),
\end{eqnarray}
where $\alpha_{\rm turb}$ is the turbulence parameter and treated as a model parameter in our study, and ${\cal Q}_{\rm cr} = 2$.
Note that $\alpha_{\rm turb}$ and $\alpha$ are defined differently in general. 
Eq. (\ref{eq:alpha}) includes the turbulence viscosity $\alpha_{\rm turb}$ and the gravitational torque, but the motion of dust induced by turbulence is considered to be related only to $\alpha_{\rm turb}$.

%
%
\subsubsection{Disk Temperature}
It is supposed that the heating sources for the disk are viscous heating and the radiation from the envelope. 
The irradiation from the central star is not taken into consideration because a sufficient amount of infalling matter is present around the disk to absorb and scatter the radiation from the central star to the disk in the disk formation stage.

The viscous heating rate per unit area of the disk is given by
$\dot E_{\rm v}=\frac{9}{4} \nu_{\rm g} \Sigma_{\rm g} \Omega^2$.
When the heating by the radiation from envelope $\sigma T_{\rm cd}^4$ and the viscous heating come into balance with the cooling by the radiation from the disk surface, the temperature of the disk surface $T_{\rm s}$ is given by
$\sigma T_{\rm s}^4=\frac{1}{2}\dot E_{\rm v}+\sigma T_{\rm cd}^4$.
Since the temperature of the disk midplane $T_{\rm mid}$ is the focus,
where the collisional growth of dust aggregates mainly takes place,
an equation that relates the disk surface temperature, $T_{\rm s}$, to the disk midplane temperature, $T_{\rm mid}$, for both optically thick and thin disks, is used. The equation is given as
\begin{eqnarray}
\label{eq:midtemp}
\sigma T_{\rm mid}^4=\frac{1}{2} \biggl(\frac{3}{8}\tau_{\rm R}+\frac{1}{2 \tau_{\rm P}} \biggl) \dot E_{\rm v}+\sigma T_{\rm cd}^4 ,
\end{eqnarray}
where $\tau_{\rm R}=\kappa_{\rm R} \Sigma _{\rm g}/2$ and $\tau_{\rm P}=\kappa_{\rm P} \Sigma_{\rm g}/2$ are the optical depths with the Rosseland mean $\kappa_{\rm R}$ and Planck mean $\kappa_{\rm P} $ opacities, respectively.
The Rosseland mean opacity used in this study is
\begin{eqnarray}
\label{eq:opacity}
\kappa_{\rm R}=
\left\{
\begin{array}{cc}
4.5 \left( \frac{T_{\rm m}}{170 \mbox{ K}} \right)^2 \ {\rm cm^2 \ g^{-1}} & (0 \mbox{ K} < T_{\rm m} <170 \mbox{ K}) \\ 4.5  \ {\rm cm^2 \ g^{-1}} & (170 \mbox{ K} < T_{\rm m} <1500 \mbox{ K}),
\end{array}
\right.
\end{eqnarray}
and the Planck mean opacity is $\kappa_{\rm P} =2.4\kappa_{\rm R}$ \citep{nakamoto1994}.
In this study, it is assumed that $T=170 \ {\rm K}$ is the evaporation temperature of ice.

\subsection{Dust Model}
\label{sec:dust}
In this study, the size distribution evolution of the dust aggregates in the disk 
is examined using the method described by \cite{brauer2008}, \cite{birnstiel2010}, and \cite{okuzumi2012}. 
Moreover, the porosity evolution of the dust aggregates is calculated.
The calculation method for the porosity evolution is similar to those
described by \cite{okuzumi2012} and \cite{kataoka2013b}.

%
%
\subsubsection{Evolution of the Dust Size Distribution}
\label{sec:secevodustdis}
When the sedimentation of dust aggregates and their turbulent stirring in the vertical direction are in equilibrium,
the vertical number density distribution of aggregates is given by a Gaussian
$({\cal N}/\sqrt{2\pi}h_{\rm d})\exp(-z^2/2h_{\rm d}^2)$,
where ${\cal N}(r,m)$ is the column number density of aggregates per unit mass at $r$ with the mass $m$,
and $h_{\rm d}(r,m)$ is the scale height of aggregates having the mass $m$.
The temporal evolution of ${\cal N}(r,m)$ is driven by the collisional growth, advection, and diffusion in the radial direction and the input from the molecular cloud core due to the infall.

The evolution of the size distribution by collisional growth is given by the vertically integrated Smoluchowski equation as \citep{birnstiel2010} 
\begin{eqnarray}
\label{eq:eq8}
\frac{\partial {\cal N}(r,m)}{\partial t}&=&\frac{1}{2} \int^m_0 K(r,m',m-m') {\cal N}(r,m') {\cal N}(r,m-m')dm'  \nonumber \\ 
                                                         &&-{\cal N}(r,m) \int^{\infty}_0 K(r,m,m') {\cal N}(r,m')dm' ,
\end{eqnarray}
where $K$ is the vertically integrated collision rate coefficient between colliding aggregates having the masses with $m_1$ and $m_2$ given by
\begin{eqnarray}
\label{eq:eq9}
K(r,m_1,m_2)=\frac{\sigma_{\rm coll}}{2 \pi h_{\rm d,1} h_{\rm d,2}} \int^\infty_{-\infty} \Delta v \exp \biggr(- \frac{z^2}{2 h_{\rm d,12}^2} \biggl) dz,
\end{eqnarray}
\normalsize
and $h_{\rm d,12}=(h_{\rm d,1}^{-2}+h_{\rm d,2}^{-2})^{-1/2}$, where $h_{\rm d,1}$ and $h_{\rm d,2}$ are the scale heights of the colliding aggregates.
Assuming perfect sticking for icy dust, the collisional cross section $\sigma_{\rm coll}$ is given by $\sigma_{\rm coll} = \pi(a_1+a_2)^2$ except when the hydrodynamic flow hinders collision between the dust aggregates \citep{sekiya2003}.

When the sedimentation and stirring of aggregates are in an equilibrium state,
the dust scale height is analytically obtained as \citep{youdin2007}  
\begin{eqnarray}
\label{eq:eq10}
h_{\rm d} = h_{\rm g} \biggl(1 + \frac{\Omega t_{\rm s}}{\alpha_{\rm turb}} \frac{1+2\Omega t_{\rm s}}{1+\Omega t_{\rm s}} \biggr)^{-1/2} ,
\end{eqnarray}
where $t_{\rm s}$ is the stopping time of the aggregates expressed as \citep{weidenschilling1977}:
\begin{eqnarray}
\label{eq:eq11}
t_{\rm s}=
\left\{
\begin{array}{cc}
 \frac{ 3m }{4 \rho_g  c_t  A} & (a<\frac{9}{4}\lambda_{\rm mfp}) \\ 
 \frac{4a}{9\lambda_{\rm mfp}} t_s^{\rm (ep)} & (a>\frac{9}{4}\lambda_{\rm mfp}, \ Re_{\rm p}<1) \\ 
 \frac{2m}{24 Re^{-0.6} \rho_{\rm g} |v_{\rm g}-v_{\rm d}| A}& (a>\frac{9}{4}\lambda_{\rm mfp}, \ 1<Re_{\rm p}<800)\\ 
 \frac{2m}{0.44 \rho_{\rm g} |v_{\rm g}-v_{\rm d}| A }& (a>\frac{9}{4}\lambda_{\rm mfp}, \ 800<Re_{\rm p}) ,
\end{array}
\right.
\end{eqnarray}
where $a$ and $A$ are the radius and projected area of a porous dust aggregate, 
$c_{\rm t}=\sqrt{8/\pi}c_{\rm s}$ is the thermal velocity,
$\lambda_{\rm mfp} = m_{\rm g}/ \sigma_{\rm moll} \rho_{\rm g}$ is the mean free path of a gas molecule,
and $\sigma_{\rm moll}= 2 \times 10^{-15} \ {\rm cm^2}$ is the collisional cross section of the gas molecules.
The relation between $a$ and $A$ is given by the same expression as Eqs. (45) - (47) in \cite{okuzumi2009}.

The particle Reynolds number of a dust aggregate, $Re_{\rm p}$, is defined as 
\begin{eqnarray}
\label{eq:prey}
Re_{\rm p} = \frac{4 a v_{\rm rel}}{\lambda_{\rm mfp}c_{\rm t} },
\end {eqnarray}
where $v_{\rm rel}$ is the relative velocity between the gas and the dust aggregate.

The relative velocity for collision of two aggregates $\Delta v$ are driven by Brownian motion, the radial and azimuthal drift motions, vertical settling, and the gas turbulence.
The relative velocity of aggregates induced by the gas turbulence is a dominant term for collision velocity, and derived analytically for Kolmogorov turbulence \citep{ormel2007}, which has three limiting cases:  
\begin{eqnarray}
\label{eq:eqdelvt}
\Delta v_t\approx
\left\{
\begin{array}{cc}
                               \delta v_{\rm g} Re_t^{1/4} \Omega |t_{\rm s,1}-t_{\rm s,2}| & (t_{\rm s,1} \ll t_\eta ) \\ 
 (1.4...1.7) \times \delta v_{\rm g} \sqrt{\Omega t_{\rm s,1}}  & (t_\eta \ll t_{\rm s,1} \ll \Omega^{-1})\\
                        \delta v_{\rm g} \biggl(\frac{1}{1+\Omega t_{\rm s,1}}+\frac{1}{1+\Omega t_{\rm s,2}} \biggr)^{1/2} &(\Omega t_{\rm s,1} \gg 1),
\end{array}
\right.
\end{eqnarray}
where $\delta v_g=\sqrt{\alpha_{\rm turb}} c_s$ is the random velocity of the largest eddies, $Re_t=D_g/\nu_{\rm mol}$ is the turbulent Reynolds number, where $D_g=\alpha_{\rm turb} c_{\rm s}^2/\Omega$ is the diffusion coefficient for the gas and $\nu_{\rm mol}= c_{\rm t} \lambda_{\rm mfp}/2$ is the molecular viscosity, $t_\eta=Re_t^{-1/2} \Omega^{-1}$ is the turnover time of the smallest eddy, and the numerical coefficient (1.4...1.7) which takes taking a numerical value of roughly 1.4--1.7, is given by the ratio of the stopping times of two colliding aggregates.

The integrand in Eq. (\ref{eq:eq9}) depends on the vertical height $z$. 
However, the dust coagulation mainly occurs at the disk midplane. 
Hence, the stopping time of the dust aggregate is evaluated at the midplane.
Then, Eq. (\ref{eq:eq9}) can be integrated analytically, and one obtains
\begin{eqnarray}
K(r,m_1,m_2)=\frac{\sigma_{\rm coll} \Delta v}{\sqrt{2 \pi} } (h_{\rm d,1}^2 + h_{\rm d,2}^2)^{-\frac{1}{2}}.
\end{eqnarray}

The evolution of the size distribution caused by advection, diffusion, and the infall is written as
\begin{eqnarray}
\label{eq:eqaddiff}
 \frac{d \Sigma_{\rm d}(r,m)}{dt} = -\frac{1}{r} \frac{\partial}{\partial r} [r (F_{\rm adv}+F_{\rm diff})]  +S_{\rm d}(r,m) ,
\end{eqnarray}
where $\Sigma_{\rm d} (r,m) = m {\cal N}(r,m)$ is the dust aggregate surface density per unit mass, 
 $F_{\rm adv}$ and $F_{\rm diff}$ are the fluxes of advection and diffusion, and $S_{\rm d}(r,m)$ is the source term of dust particles.
The advection flux is given by
$ F_{\rm adv} = v_{\rm r}(r,m) \Sigma_{\rm d}(r,m)$, 
where $v_{\rm r}$ is the velocity of dust aggregates in the radial direction.
The diffusion flux is written as
\begin{eqnarray}
\label{eq:eqfdiff}
F_{\rm diff} &=& -D_{\rm d}(r,m) \frac{\partial}{\partial r} \biggl(\frac{\Sigma_{\rm d}(r,m)}{\Sigma_{\rm g}}  \biggr)  \Sigma_{\rm g},
\end{eqnarray}
where $D_{\rm d}$ is the diffusion coefficient for dust.

The velocity of aggregates in the radial direction, $v_{\rm r}(r,m)$, is given by 
\begin{eqnarray}
\label{eq:eq12}
v_{\rm r}=-\frac{\Omega t_{\rm s}}{1+ (\Omega t_{\rm s})^2} 2\eta v_{\rm k}+\frac{v_{\rm g}}{1+(\Omega t_{\rm s})^2},
\end{eqnarray}
where $2\eta$ is the ratio of the pressure gradient force to the stellar gravity force in the radial direction, and $\eta$ is given by
$\eta= -\frac{1}{2}\left(\frac{c_{\rm s}}{v_{\rm K}} \right)^2
\frac{\partial \ln(\rho_{\rm g}c_{\rm s}^2)}{\partial \ln r}$, 
where $v_{\rm K}=r \Omega$ is the Kepler velocity.
Note that the order of $\eta$ is determined by $\left(\frac{c_{\rm s}}{v_{\rm K}} \right)^2$ since $\frac{\partial \ln(\rho_{\rm g}c_{\rm s}^2)}{\partial \ln r} \sim {\cal O}(1)$.
The first term on the right-hand side of Eq. (\ref{eq:eq12}) expresses the radial drift velocity caused by the disk gas,
which has sub-Keplerian motion, and its absolute value has the maximum $\eta v_{\rm K}$ when $\Omega t_{\rm s}=1$.
The second term represents the motion induced by the radial flow of the disk gas.
The diffusion coefficient for the dust is given by $D_{\rm d}= D_{\rm g}/[1+(\Omega t_{\rm s})^2]$  \citep{youdin2007}, where $D_{\rm g}$ is the diffusion coefficient for the gas.
It is assumed that $D_{\rm g}$ is equal to the turbulent gas viscosity $\nu_{\rm g}$.
The third term in Eq. (\ref{eq:eqaddiff}) shows the source term of dust including the infall of dust from the molecular cloud core and the condensation of water vapor that originates from the snowline. 
The details of the source term will be described in the next section.

In this study, the evolution of the dust aggregate volume $V(r,m)$ is also considered using the method described by \cite{okuzumi2009}.
In this method, the temporal evolution of the quantity $V{\cal N}$ is calculated, and the average volume of the dust aggregate at each orbital radius $r$ with aggregate mass $m$ is obtained. 
Its collisional term is given by 
\begin{eqnarray}
\label{eq:eq14}
\frac{\partial (V{\cal N})}{\partial t}&=&\frac{1}{2} \int^m_0 [V_{1+2}K](r,m',m-m') {\cal N}(r,m') {\cal N}(r,m-m')dm'  \nonumber \\ 
                                                        &&-V(r,m){\cal N}(r,m) \int^{\infty}_0 K(r,m,m') {\cal N}(r,m')dm' .
\end{eqnarray}
The function $[V_{1+2}K](r, m_1, m_2)$ is written as
\begin{equation}
\label{eq:eqvkernel}
[V_{1+2}K](r,m_1,m_2)=\frac{\sigma_{\rm coll} \Delta v V_{1+2}}{\sqrt{2 \pi} } (h_{\rm d,1}^2 + h_{\rm d,2}^2)^{-\frac{1}{2}}, 
\end{equation}
and $V_{1+2}$ is the volume of merged aggregates. 
The function of $V_{1+2}$ will be given in Section \ref{sec:secvevo}.
The evolution of $(V{\cal N}) (r,m)$ by advection, diffusion, and the source is calculated in the same manner as ${\cal N}(r,m)$.

%
\subsubsection{Source Term of Dust}
The second term on the right-hand side of Eq. (\ref{eq:eqaddiff}), the source term, includes the mass accretion from the molecular cloud core and the condensation of icy dust from the water vapor that is supplied from inside the snowline. 
The size of the infall dust is assumed to be $a_0=0.1 \ {\rm \mu m}$, which is the monomer size in our calculations.
Assuming that the dust-to-gas mass ratio is 0.01 in the molecular cloud core,
the source term due to the infall is given by
$S_{\rm d, infall}=\delta[m-m_0] \times 0.01 S_{\rm g}$.

The condensation of icy dust particles close to the snowline is also the mechanism that prompts the increase in the dust surface density and the growth of dust particles \citep{ros2013}.
However, the size of the condensed particles is uncertain because it depends on the cooling rate, pressure, and so on.
The size of the condensed monomer particle may influence the collisional growth of porous dust aggregates \citep{arakawa2016}.
In this study, however, to simplify the calculation, the size of condensed monomers is assumed to be the same as
the size of the infall dust. 
We calculate the mass flux of water vapor across the snowline by advection and diffusion, and we assume that the vapor condenses as icy monomers.
For example, if the snowline migrates inward, all the water vapor in the region that was $T > 170$ K condenses as icy monomers. 
Then, the abundance of condensed icy dust near the snowline is calculated and added to the source term in Eq. (\ref{eq:eqaddiff}).

%
\subsubsection{Porosity Change}
\label{sec:secvevo}
The evolution of the aggregate porosity, i.e., $V_{1+2}$ in Eq. (\ref{eq:eq14}), is taken into consideration.
The collisional compression of aggregates depends on the rolling energy between two contacting monomers $E_{\rm roll}$
and the impact energy of the two aggregates $E_{\rm imp}=m_1 m_2 \Delta v^2/2(m_1+m_2)$.
When $E_{\rm imp} \ll E_{\rm roll}$, collisional compression is not effective.
This case is called a hit-and-stick collision, and the volume of the aggregate after collision is given by
$V_{1+2} =V_1 +V_2 + V_{\rm void}   \ \ \ (E_{\rm imp} \ll E_{\rm roll})$,
where
$V_{\rm void} ={\rm min} \biggl\{0.99-1.03 \ln \biggl(\frac{2}{V_1/V_2+1}  \biggr), 6.94 \biggr\} V_2$
is the volume of the void formed after the collision of two aggregates \citep{okuzumi2009}.

In contrast, when $E_{\rm imp} \gg E_{\rm roll}$, collisional compression becomes effective.
In this case, the porosity of aggregates no longer increases owing to the compression of the void by collision, and the internal density of the aggregate remains nearly constant with collisional growth (Sec 3.2.2 in \cite{okuzumi2012}).
Thus, the volume evolution is given by a simple equation:
\begin{equation}
\label{eq:eqvevo2}
V_{1+2} =V_1 +V_2  \ \ \ (E_{\rm imp} \gg E_{\rm roll}). 
\end{equation}
Note that the formula for the volume evolution that takes collisional compression into account should actually depend on the two volumes of colliding aggregates and the impact energy. In fact, the recipe obtained from numerical experiments is a function of the impact energy \citep{suyama2012}.
However, these numerical experiments only examined collisions between aggregates having similar sizes,
and there is no recipe for different-sized collisions.
In the present study, as will be shown later, both similar-sized and different-sized collisions need to be considered. 
Thus, the volume evolution by collisional compression is assumed to be described by the simple expression shown here.
The effect of this volume evolution on the results will be discussed later.

Aggregates also suffer the static compression by the gas pressure and self-gravity.
\cite{kataoka2013a} investigated the strength of highly porous aggregates against static compression and gave the compressive strength of the aggregates, $P$, as
$P=\frac{E_{\rm roll}}{a_0^3} \biggl( \frac{\rho_{\rm int}}{\rho_0} \biggr)^3$.
When the aggregate suffers a pressure higher than the compressive strength, the aggregate is compressed until its strength becomes equal to the static pressure.
The volume of a dust aggregate of which the compressive strength equals the pressure $P$ is given by 
\begin{equation}
V=\left( \frac{a_0^3}{E_{\rm roll}} P \right) ^{-1/3} \frac{m}{\rho_0}.
\label{eq:internaldensity}
\end{equation}
Further, the static compression due to the gas pressure $P_{\rm gas}$ and the pressure caused by the self-gravitational force  $P_{\rm grav}$ are given by
$P_{\rm gas}=\frac{m v_{\rm rel}}{\pi r^2} \frac{1}{t_{\rm s}}$
and
$P_{\rm grav}=\frac{Gm^2}{\pi r^4}$, respectively.

%
%
\subsection{Numerical Method}
In this study, Eqs. (\ref{eq:eq1}), (\ref{eq:eq8}), and (\ref{eq:eq14}) are solved numerically with an explicit time-integration scheme.
The advection terms for the gas and dust are calculated by a first-order upwind scheme.
The inner and outer boundaries are set to not influence the region where the icy dust particles are present
for each model parameter.

The dust coagulation terms are calculated using the method given by \cite{okuzumi2009}. 
At the center of each radial cell, the bins of the dust aggregate mass are set as
  $m_{\rm k}=k m_0$ for $k \leq N_{\rm bd}$ and
  $m_{\rm k}=m_{\rm k-1} 10 ^{1/N_{\rm bd}}$ for $k \geq N_{\rm bd}+1$,
where
$m_0$ is the monomer particle mass,
$k$ and $N_{\rm bd}$ are positive integers,
and $N_{\rm bd}=40$, as used by \cite{okuzumi2012}. 
The time increment $\Delta t $ is decided at every time step
so that the fractional decreases in ${\cal N}$ and $V{\cal N}$ remain lower than 0.5 at all bins.

\section{RESULTS}
\label{sec:result}

%
%
\subsection{Steady Disk Model}
First, the results of the steady disk model are shown to compare it with the disk formation and evolution models, which will be shown later. 
Moreover, it will be shown that our results using the coagulation equation well-reproduce the results obtained analytically \citep{kataoka2013b}.
The minimum-mass solar nebula (MMSN) model \citep{hayashi1981} for the gas radial distribution with a central star having a solar mass, is employed.
Thus, the gas surface density $\Sigma_{\rm g}$ is given by $\Sigma(r)=1700(r/1 {\rm AU}) \ {\rm g \ cm^{-2}}$.
The disk midplane temperature $T$ is given by $T(r)=137(r/1 {\rm AU}) \ {\rm K}$ \citep{chiang2001}.
The initial dust-to-gas mass ratio is assumed to be 0.01, and the initial size of all dust particles is
set to be $a_0=0.1 \ {\rm \mu m}$, which is the monomer size.
This disk model corresponds to the model taken by \cite{kataoka2013b}.

\begin{figure}[h] 
  \centering
  \includegraphics[width=0.45\textwidth]{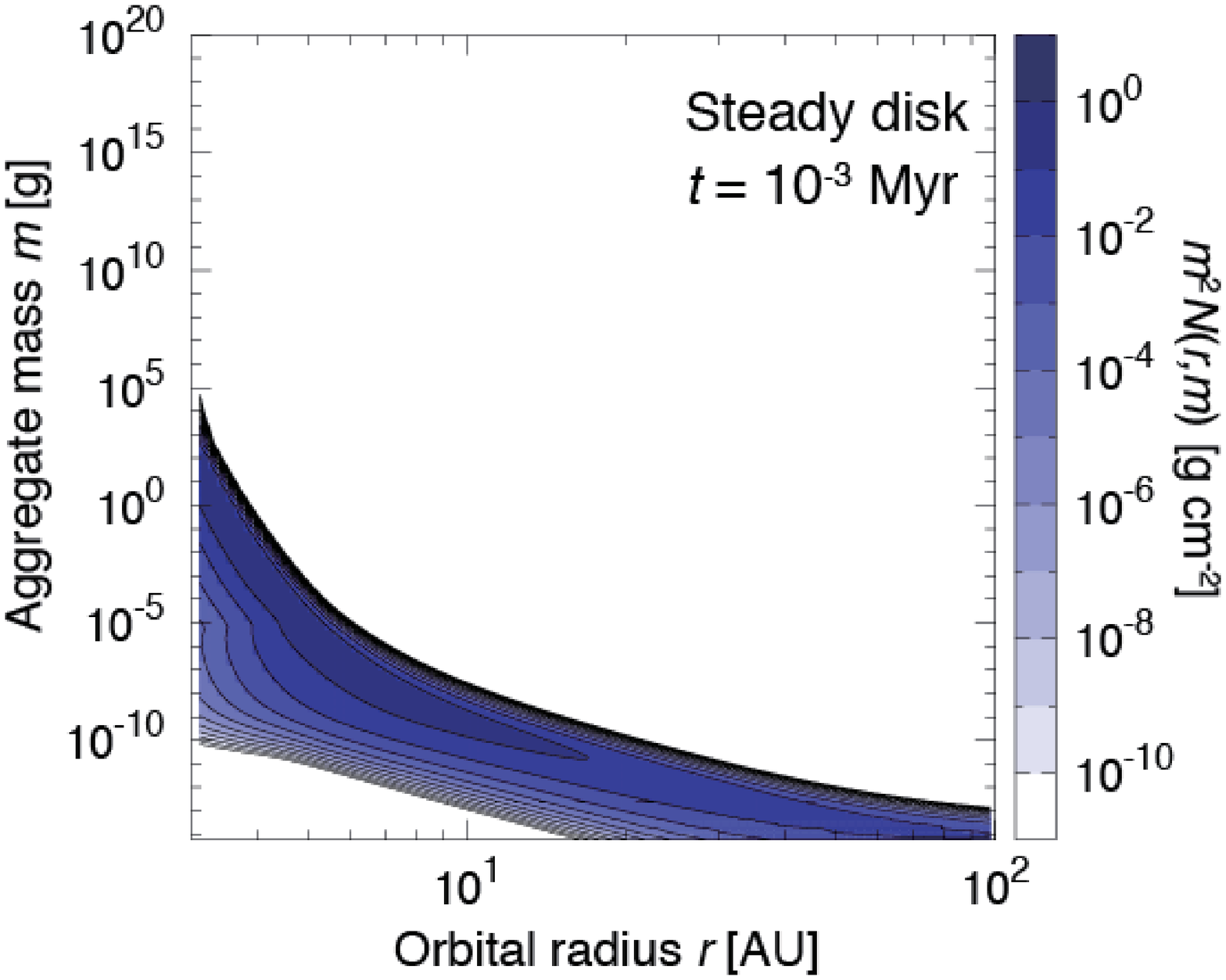}

   \includegraphics[width=0.45\textwidth]{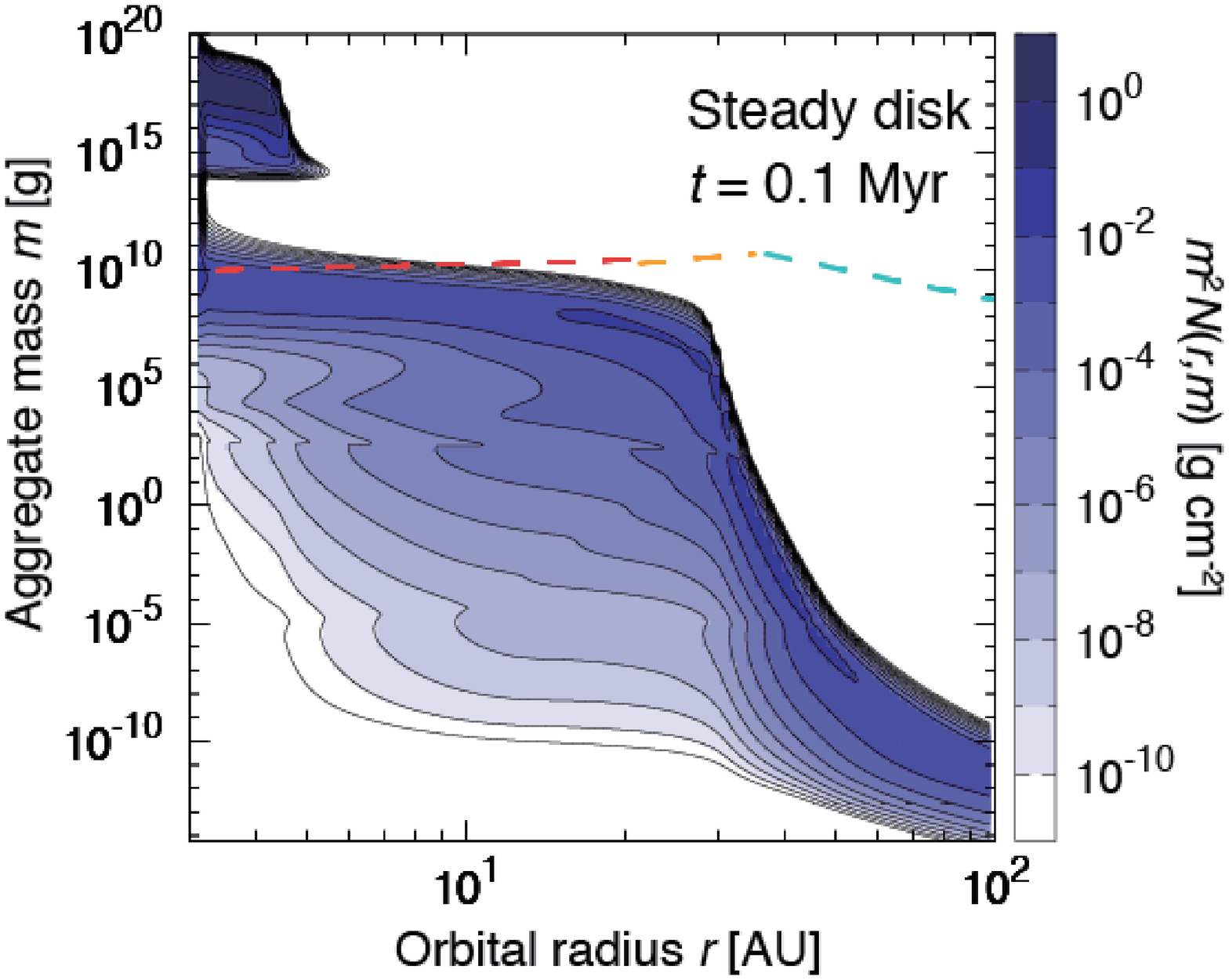}
  \caption{Aggregate size distribution $m^2 {\cal N}$ at $t= 10^{-3}$ and $0.1 \ {\rm Myr}$ for the steady disk model as a function of the orbital radius $r$ and aggregate mass $m$. The colored dashed curve shows the size corresponding to $\Omega t_{\rm s}=1$. Red, yellow, and blue lines indicate Allen, Stokes, and Epstein's laws, respectively.}
  \label{fig:fig1}
\end{figure}

Figure \ref{fig:fig1} shows the radial size distribution at $t=10^{-3} $ and $0.1\  {\rm Myr}$ for the steady disk model.
Aggregates grow to a larger size than the size of the radial drift barrier at
$\Omega t_{s}=1$ (dashed curve in Figure \ref{fig:fig1}) in the inner region of the disk ($r < 5 \ {\rm AU}$).
In the middle region ($5 \ {\rm AU} < r < 30 \ {\rm AU}$), aggregates drift inward,  
while in the outer region ($30 \ {\rm AU} < r$)
aggregates do not drift considerably.
This is because aggregates in the outer region
are small and the Stokes number of them is much smaller than unity.
In the outer region, the dust growth timescale becomes long
so  0.1 Myr is not enough for aggregates to grow.

\begin{figure}[t]
  \centering
  \includegraphics[width=0.45\textwidth]{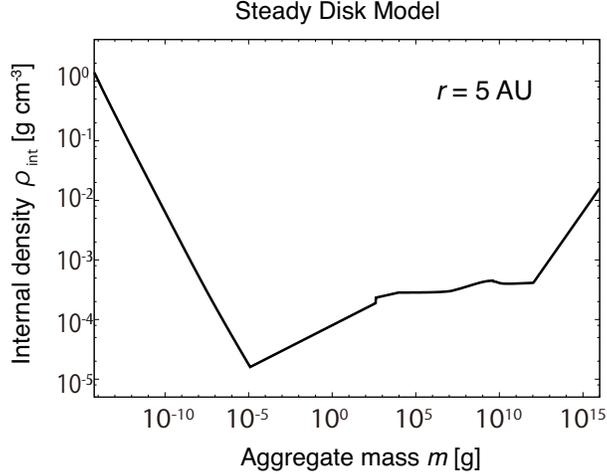}
  \caption{Evolution of the internal density $\rho_{\rm int} $ at $r=5 {\rm AU}$ for the steady disk model as a function of the aggregate mass $m$. }
  \label{fig:fig2}
  \end{figure}

Figure \ref{fig:fig2} shows the evolution of the aggeregate internal density  $\rho_{\rm int} $ at $r=5 \ {\rm AU}$.
When the aggregate mass is small ($m < 10^{-5} \ {\rm g}$), the internal density evolution is almost
equal to that of fractal aggregates with the fractal dimension $d_{\rm f} \simeq 2$
because the aggregates grow mainly through collisions with similarly sized aggregates. 
For larger sizes ($10^{-5} \ {\rm g} < m < 10^{12} \ {\rm g}$), gas compression becomes effective, and the aggregate internal density increases with the mass in accordance with the equations for $V$ and $P_{\rm gas}$.
In much larger size ranges ($10^{12} \ {\rm g}< m$), self-gravitational compression becomes effective, as described by the equations for $V$  and $P_{\rm grav}$. 
It is noted that these results are consistent with the results by \cite{kataoka2013b}, who investigated the growth and radial drift of dust aggregates in the same gas disk used in our steady disk model.

\begin{figure}[t]
  \centering
  \includegraphics[width=0.45\textwidth]{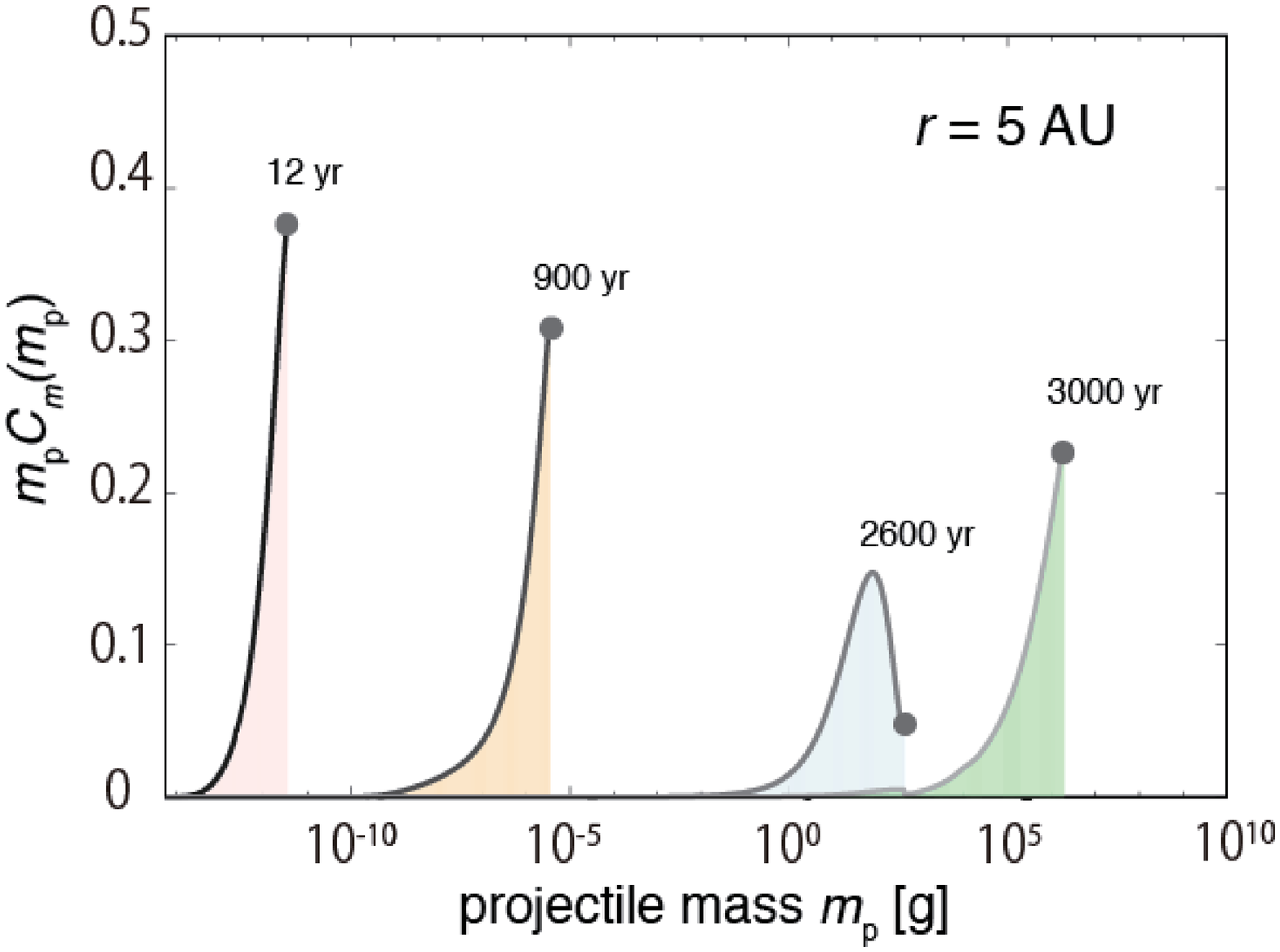}
  \caption{Projectile mass distribution per unit logarithmic projectile mass: $m_{\rm p}C_{m}(m_{\rm p})$ for different targets with mass $\left<m \right >_m$. The circles show the points of equally sized aggregation (i.e., $m_{\rm p} = m$). }
  \label{fig:figs-prodis}
\end{figure}

It is important to determine the size during collision that contributes to the growth of the aggregate the most
because the size of the  (projectile) aggregate with the highest contribution influences
the porosity and growth rate of the (target) aggregate.
To see the contribution, the projectile mass distribution function \citep{okuzumi2009} is defined as follows:
\begin{eqnarray}
C_{m}(m_{\rm p})=\frac{m_{\rm p} K(m_{\rm p}, m) {\cal N}(m_{\rm p})}{\int^m_{m_0} m_{\rm p}' K(m_{\rm p}', m) {\cal N}(m_{\rm p}') dm_{\rm p}' },
\end{eqnarray}
where $m_{\rm p} < m$ is the projectile aggregate mass, $m$ is the target aggregate mass, and $m_{\rm p} < m$.
Figure \ref{fig:figs-prodis} shows the projectile mass distribution per unit $\ln m_{\rm p}$ for different targets with mass $ \left< m \right>_m$ at $r=5 \ {\rm AU}$ for the steady disk model. 
The weighted average mass $ \left<m \right >_m$ is defined by 
\begin{eqnarray}
\left<m \right>_m = \frac{\int m^2 {\cal N} dm}{\int m {\cal N} dm} .
\end{eqnarray}
The weighted average mass approximately corresponds to the aggregate mass at the peak of the mass distribution (see, e.g., \cite{okuzumi2012}).
In Figure \ref{fig:figs-prodis}, it is seen that the growth of the target with mass $\left<m \right >_m$ is dominated by projectiles with a similar mass as the target for each target size.
For $m<10^{-5} \ {\rm g}$, this similarly sized aggregation results in the high porosity evolution with $d_{\rm f} \simeq 2$ (Figure \ref{fig:fig2}).

\subsection{Evolutionary Disk: Fiducial Model}
\label{sec:resultedisk}
Next, the results of the evolutionary disk models described in Section \ref{sec:gas} are presented and how disk evolution affects the growth of icy dust aggregates is shown.
We calculated with four different parameters, and the model parameters are summarized in Table \ref{tab:tab1}.

\begin{table}[h]
  \begin{center}
    \caption{Model parameters in our study}
    \begin{tabular}{|c|| c| c |c|} \hline
      Model &$\omega_{\rm cd}$ & $\alpha_{\rm turb} $& $T_{\rm cd} $ \\ \hline 
      Typical values &$10^{-15} \ - 10^{-13} \ {\rm s^{-1}} \ ^{\rm [1]}$ & $10^{-5} \ - 10^{-1} \ ^{\rm [2]}$ & 10 - 20 K  $^{\rm [3]}$ \\ \hline \hline
      Fiducial &$2 \times 10^{-14} \ {\rm s^{-1}}$ & $10^{-3}$ & 15 K  \\ \hline
      A &$2 \times 10^{-14} \ {\rm s^{-1}}$ & $10^{-4}$ & 15 K \\ \hline
      B &$2 \times 10^{-14} \ {\rm s^{-1}}$ & $10^{-3}$ & 20 K \\ \hline  
       C &$5 \times 10^{-15} \ {\rm s^{-1}}$ & $10^{-3}$ & 15 K \\ \hline 
      \end{tabular}
    \label{tab:tab1}
    
    {\scriptsize [1] \cite{goodman1993}, [2] \cite{van1993}, [3]  \cite{hartmann1998}}
  \end{center}
\end{table}

\subsubsection{Gas Disk Evolution} 
\label{sec:secgasresult}

\begin{figure}[h]
\centering
  \includegraphics[width=0.45\textwidth]{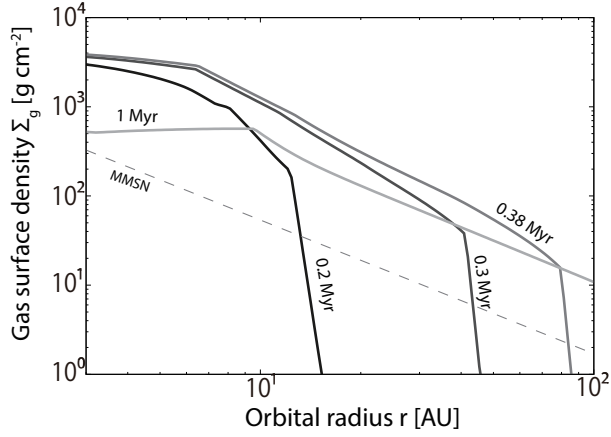}
  \caption{Gas surface density $\Sigma_{\rm g }$ at different times for the fiducial model as a function of the orbital radius $r$ (solid curves). The dashed line shows that of the MMSN model.}
  \label{fig:fig3}
  \end{figure}

The gas surface density evolution of the fiducial model is displayed in Figure \ref{fig:fig3}. 
The gas surface density at each orbital radius is an increasing function of the time during the infall stage
when the mass accretion from the molecular cloud core continues ($t< 0.38 \ {\rm Myr}$),
while it decreases after the mass accretion from the molecular cloud core ceases ($0.38 \ {\rm Myr} < t < 1 \ {\rm Myr}$) because of the diffusive mass flow in the radial direction in the disk.
Note that once the gas surface density becomes very high, the gas disk undergoes gravitational instability;
then, the gas surface density does not increase further owing to the angular momentum transport by the gravitational torque even if the mass accretion from the molecular cloud core continues.
Figure \ref{fig:figQ} shows Toomre's ${\cal Q}$ value \citep{toomre1964} at different times,
which is the measure of the gravitational stability of the gas disk.
It is seen that the outer region (6-90 AU) of the disk becomes gravitationally unstable at 0.38 Myr.  

\begin{figure}[h]
\centering
 \includegraphics[width=0.45\textwidth]{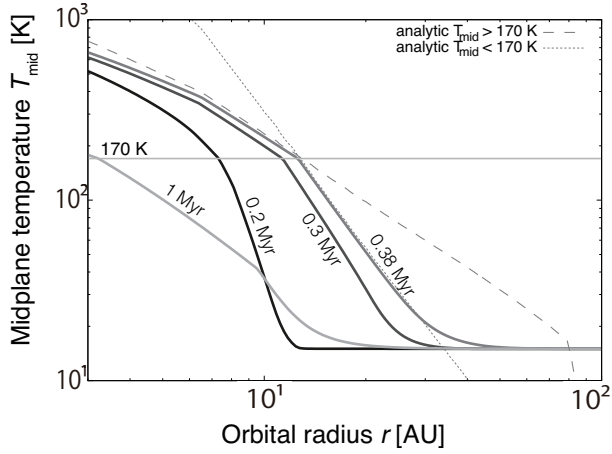}
  \caption{Disk temperature $T_{\rm mid}$ at the disk midplane at each time as a function of the orbital radius $r$ (solid curves).  The horizontal solid line shows the ice evaporation temperature $T= 170 \ {\rm K}$. The gray dashed and dotted lines show the analytical solutions of the midplane temperature at 0.38 Myr with ${\dot M}_{\rm star}=10^{-6} \  M_{\rm \odot} \ {\rm yr^{-1}}$.}
  \label{fig:figatemp}
\end{figure}

Figure \ref{fig:figatemp} shows the disk midplane temperature at different times. 
The dominant heating source for the disk is viscous heating at the disk midplane; thus, the temperature is an increasing function of the surface gas density. 
After the infall stage, the midplane temparature decreases with the time, and the snowline migrates to 3 AU at 1 Myr.
The disk temperature influences the viscous evolution of the gas disk, the collision rate of the aggregates,
and so on.
In particuar, the location of the snowline, which is mainly determined by the disk temperature,
plays an important role in the growth of icy dust aggregates. 
Outside the snowline, icy dust aggregates can be present and grow by mutual collisions.
In contrast, inside the snowline, H$_2$O molecules exist as water vapor, and no icy solid particles would be present.
It is seen that the snowline reaches about 12 AU at 0.38 Myr, and this is the maximum radius of the snowline location because no material falls from the molecular cloud core after this.

The gray dashed and dotted lines in Figure \ref{fig:figatemp} show the analytical solutions of the midplane temperature for $T<170 \ {\rm K}$ and $T>170 \  {\rm K}$, respectively, which are derived as follows. 
The heating rate per unit area with the steady accretion rate ${\dot M} = 2 \pi r v_{g, r} \Sigma_{\rm g}$ is given by Eq. (\ref{eq:eq2}) and $\dot E_{\rm v}=\frac{9}{4} \nu_{\rm g} \Sigma_{\rm g} \Omega^2= {\rm constant}$ as  
\begin{eqnarray}
\label{eq:heatrate}
\dot E_{\rm v}=\frac{3GM{\dot M}}{4 \pi r^3}. 
\end{eqnarray}
In the optically thick region, by the assumption that $\tau_{\rm R} \gg 1$, the midplane temperature is given by Eqs. (\ref{eq:midtemp}) and (\ref{eq:heatrate}) as 
\begin{eqnarray}
\label{eq:anatemp}
\sigma T_{\rm mid}^4=\frac{9GM{\dot M}}{128 \pi r^3} \kappa_{\rm R} \Sigma_{\rm g}. 
\end{eqnarray}
Using the opacity by Eq. (\ref{eq:opacity}), the midplane temperature is approximately given as

\begin{eqnarray}
\label{eq:anatemp2}
T_{\rm mid} \simeq 
170 \biggl(\frac{r}{14 \ {\rm AU}} \biggr)^{-3q} 
\biggl(\frac{{\dot M}_{\rm star}}{10^{-6} \ M_{\odot} \ {\rm yr^{-1}}} \biggr)^{q} 
\biggl(\frac{\Sigma_{\rm g}}{10^{3} \ {\rm g \ cm^{-2}}} \biggr)^{q} 
\biggl(\frac{M}{M_{\odot}} \biggr)^{q} \mbox{ K}, 
\end{eqnarray}
where the constant value $q$ is given as $q =\frac{1}{2} \ (0 \mbox{ K} < T_{\rm m} <170 \mbox{ K})$ or $q=\frac{1}{4} \ (170 \mbox{ K} < T_{\rm m} <1500 \mbox{ K)}$.

\begin{figure}[h]
\centering
 \includegraphics[width=0.45\textwidth]{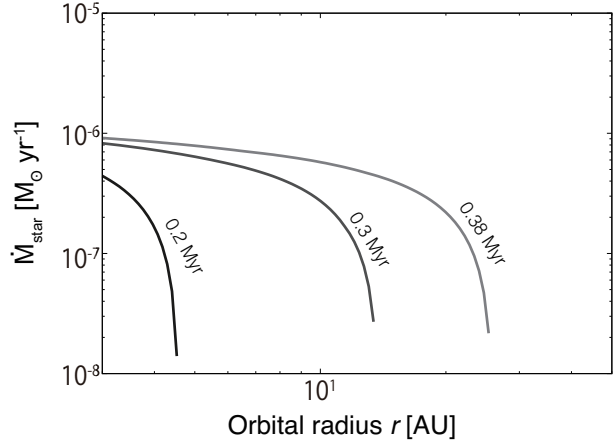}
  \caption{Mass accretion rate in the disk toward the central star ${\dot M}_{\rm star} = -2 \pi r v_{r} \Sigma_{\rm g}$ at each time.}
  \label{fig:mdot}
\end{figure}

\begin{figure}[h]
  \centering
    \includegraphics[width=0.45\textwidth]{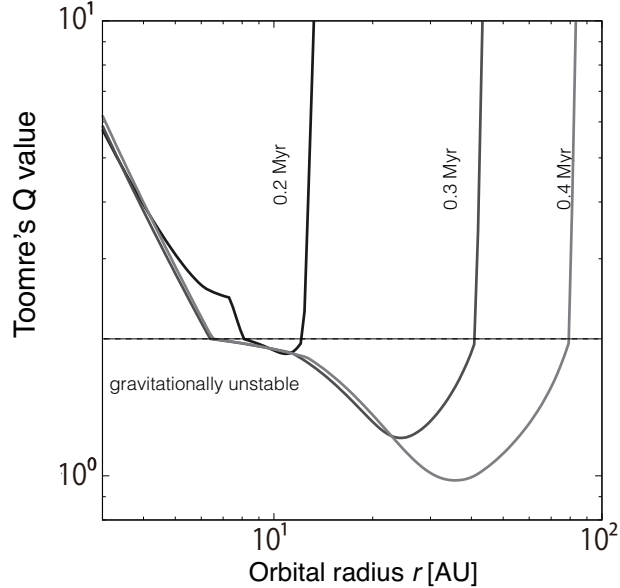}
  \caption{Toomre's ${\cal Q}$ value \citep{toomre1964} at different times for the fiducial model as a function of the orbital radius $r$ (solid curves). The dashed line shows ${\cal Q}_{\rm cr} = 2 $.}
  \label{fig:figQ}
\end{figure}

Figure \ref{fig:mdot} shows the mass accretion rate of the gas in the disk toward the central star ${\dot M}_{\rm star} = -2 \pi r v_{r} \Sigma_{\rm g}$ at different times. 
It seems that steady accretion is achieved inside 10 AU at 0.38 Myr with the accretion rate ${\dot M}_{\rm star}\sim 10^{-6} \ M_{\odot} \ {\rm yr^{-1}}$.
By using the appropriate ${\dot M}_{\rm star}$, we can see that the analytical solution well-reproduces the numerical results.

Toomre's ${\cal Q}$ value is displayed in Figure \ref{fig:figQ}. 
This shows that the gas disk is gravitationally unstable in the outer region ($r>6 \ {\rm AU}$). 
In this region, large-scale angular momentum transport occurs and value of the gas surface density reaches upper limit.

\begin{figure}[h]
  \centering
  \includegraphics[width=0.45\textwidth]{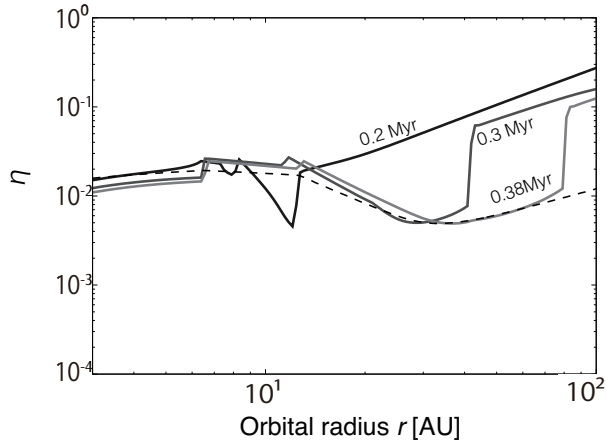}
  \caption{The ratio of the pressure gradient to the gravity $\eta$ at each time as a function of the orbital radius $r$(solid curves). The dashed curve shows $2\times c_{\rm s}^2/v_{\rm K}^2$ at 0.38 Myr.}
  \label{fig:figaeta}
\end{figure}

Figure \ref{fig:figaeta} shows $\eta$, which is related to the ratio of the pressure gradient force to the stellar gravity force in the radial direction. 
The value of $\eta$ influences the radial drift velocity of the aggregate.
The dashed curve in Figure \ref{fig:figaeta} shows $2\times c_{\rm s}^2/v_{\rm K}^2$ at 0.38 Myr.
We can see that the order of $\eta $ is determined by the square of the ratio of the sound speed at each orbit to Kepler's  velocity, except for areas where the spatial density gradient is steep.
The growth conditions for aggregates using $\eta$ will be discussed in Section \ref{sec:secgcon}.

\subsubsection{Dust Disk Evolution} 
\begin{figure}[h]  
\centering
  \includegraphics[width=0.45\textwidth]{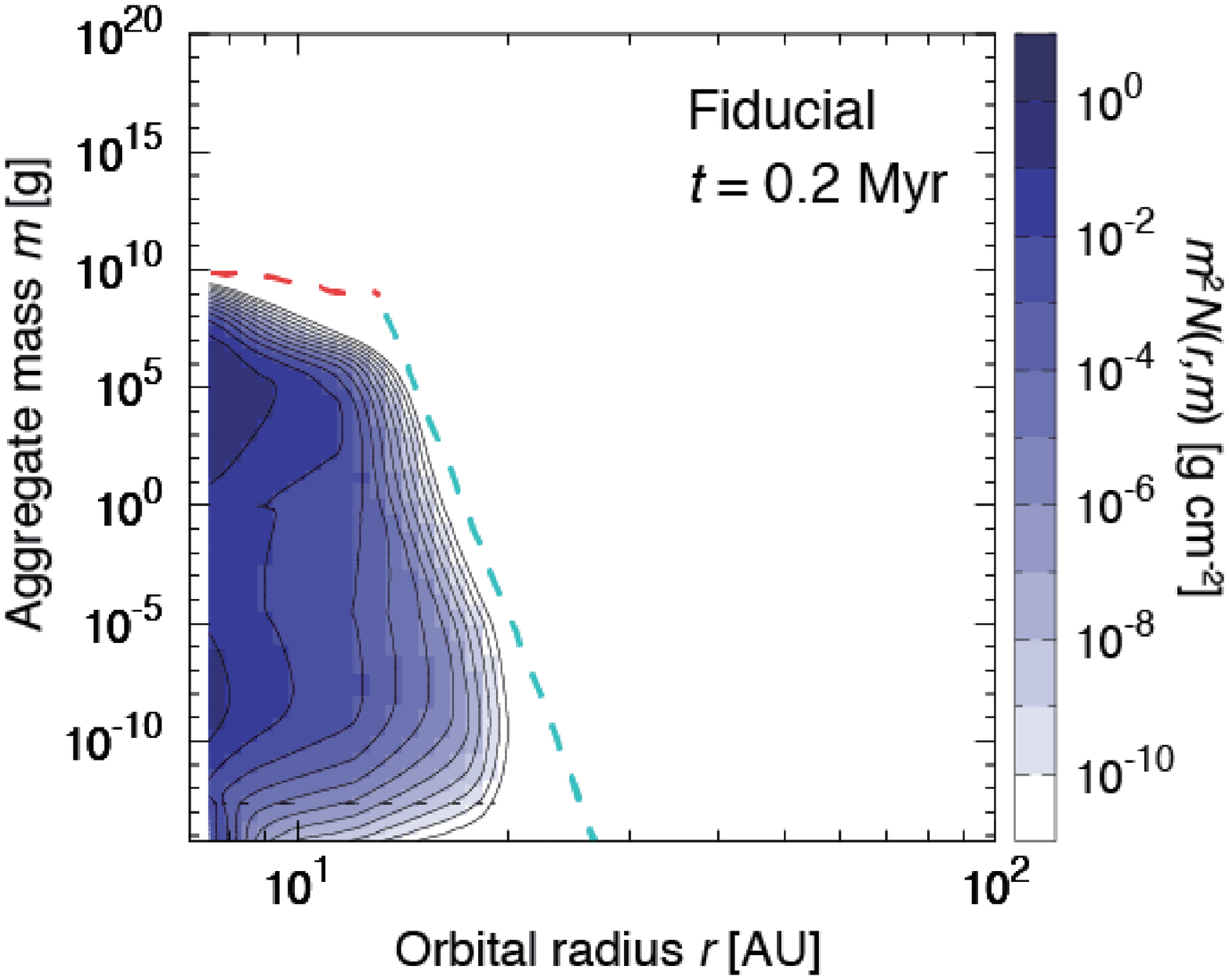}
    \includegraphics[width=0.45\textwidth]{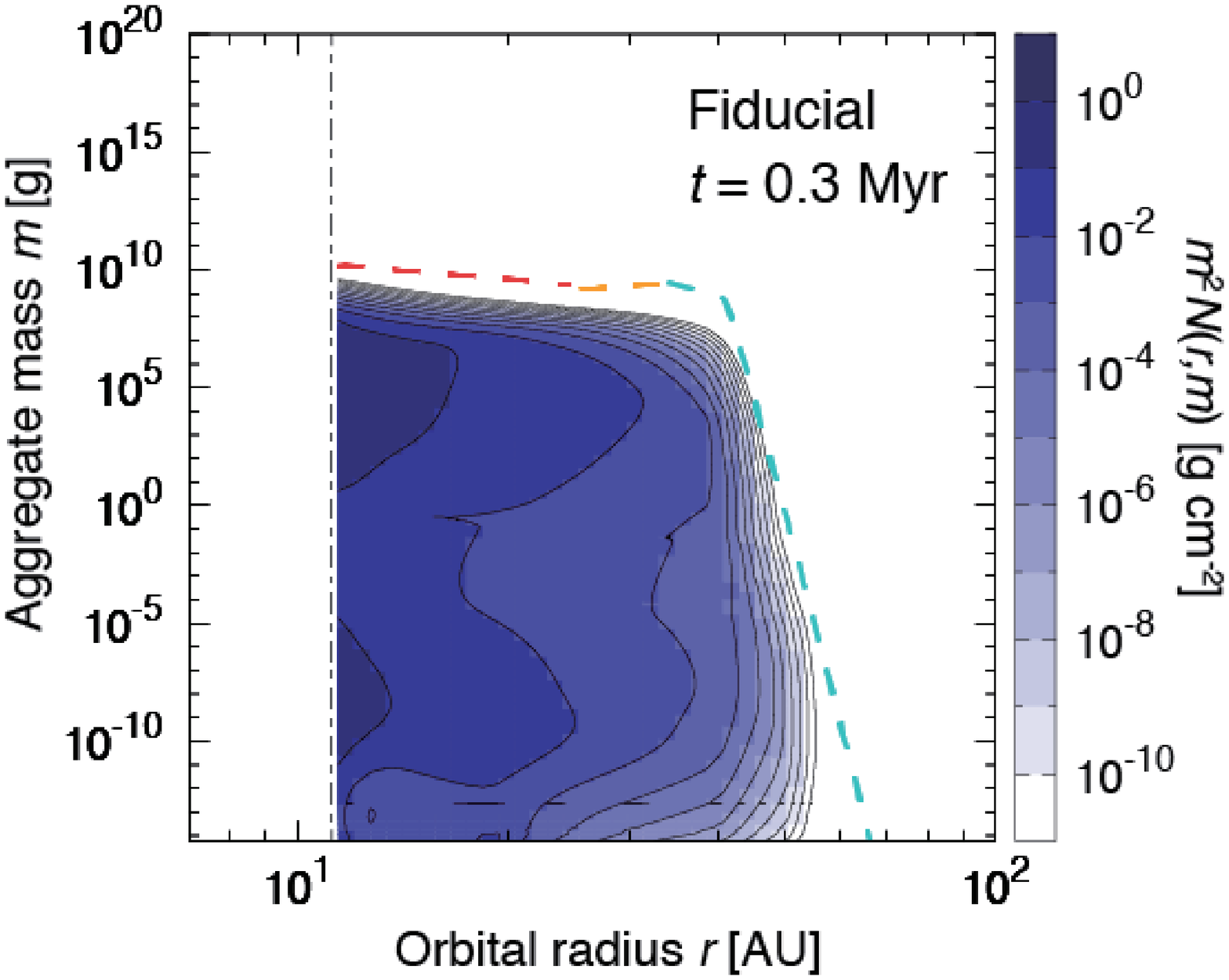}
 \includegraphics[width=0.45\textwidth]{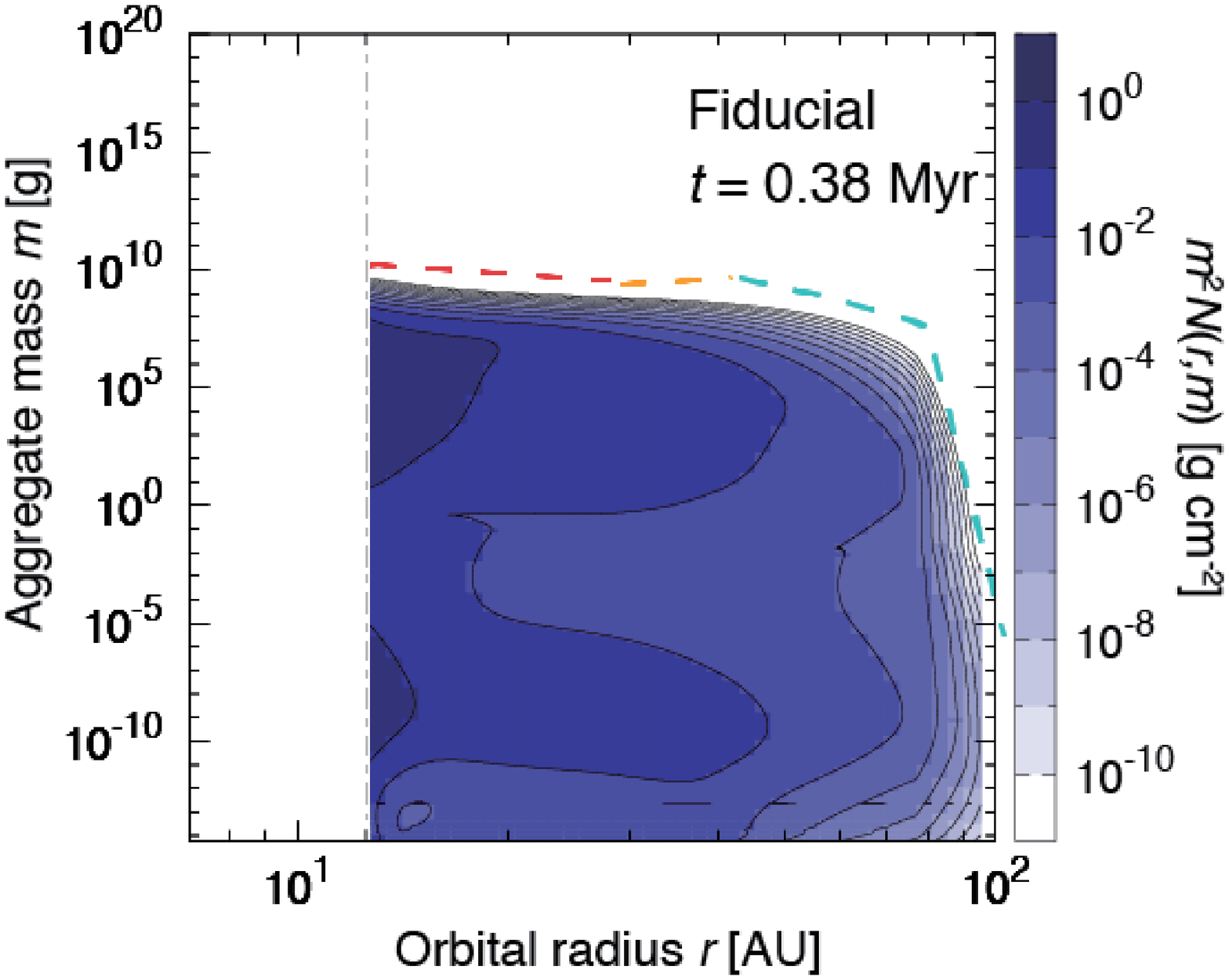}
 \caption{Aggregate size distribution $m^2 {\cal N}$ at different times for the fiducial model as a function of the orbital radius (from 7 AU to 100 AU) $r$ and aggregate mass $m$. The vertical dashed line shows the snowline. The colored dashed line shows the size corresponding to $\Omega t_{\rm s}=1$. The red, yellow, and blue lines indicates Allen's, Stokes's, and Epstein's laws, respectively, at $\Omega t_{\rm s}=1$.}
   \label{fig:fig4}
\end{figure}

The evolution of icy dust aggregates for the fiducial model is shown here.
Figure \ref{fig:fig4} shows snapshots of the size distribution of the aggregates at different times.
At each radius in the disk, as the aggregates reach $\Omega t_{\rm s} \simeq 1$,
the radial drift overcomes the growth; therefore, no aggregate exceeds the size corresponding to $\Omega t_{\rm s} \simeq 1$.

\begin{figure}[h] 
  \centering
  \includegraphics[width=0.45\textwidth]{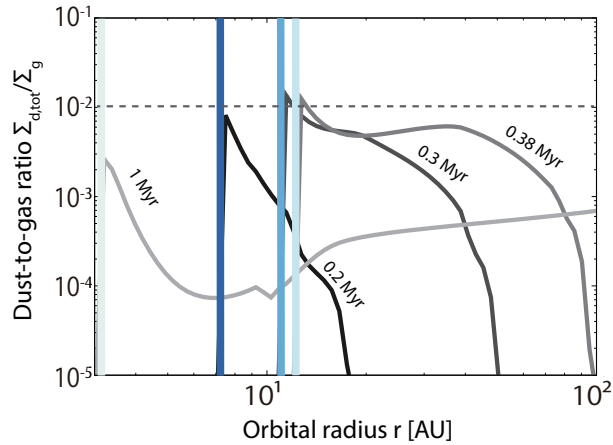}
  \caption{Dust-to-gas ratio $\Sigma_{\rm d,tot}/\Sigma_{\rm g}$ at different times as a function of the orbital radius (gray curves). The blue vertical lines show the orbital radii of the snowline at each time. The horizontal dotted line show the initial dust-to-gas mass ratio in the molecular cloud core.}
  \label{fig:figdsden}
\end{figure}

Figure \ref{fig:figdsden} shows the dust-to-gas mass ratio $\Sigma_{\rm d,tot}/\Sigma_{\rm g}$ at different times as a function of orbital radius $r$, where $\Sigma_{\rm d,tot} = \int \Sigma_{\rm d}(r,m) dm$ is the dust surface density.
Just outside the snowline, there is a pile-up of icy aggregates. This is caused by the inward radial drift of icy dust aggregates outside the snowline and the newly formed icy monomers caused by condensation of water vapor coming from inside the snowline due to diffusion.
On the other hand, it can be seen that the dust-to-gas ratio decreases in the outer part of the disk because the gas spreads outward by viscous evolution, while the dust aggregates drift toward the center star.
A similar effect was described by \cite{birnstiel2014}.
This decrease of the dust-to-gas mass ratio is caused because the radial drift timescale of dust aggregates is much shorter than the gas flow timescale in the disk.

Our numerical results show a contrast to  observational results by \cite{ansdell2016}.
According to their observations, young disks (estimated ages are $1-3$ Myr old) are enriched in dusts relative to  the interstellar medium.
We speculate that the dust-to-gas mass ratio would be caused by some mechanisms including the photoevaporation, the disk wind, and so forth, in addition to the dust growth and the radial drift.
Those gas dispersal effects may increase the dust-to-gas mass ratio in the disk.
Since those effects are not taken into account in out current study, they should be examined in the future.

\begin{figure}[h] 
  \centering
  \includegraphics[width=0.45\textwidth]{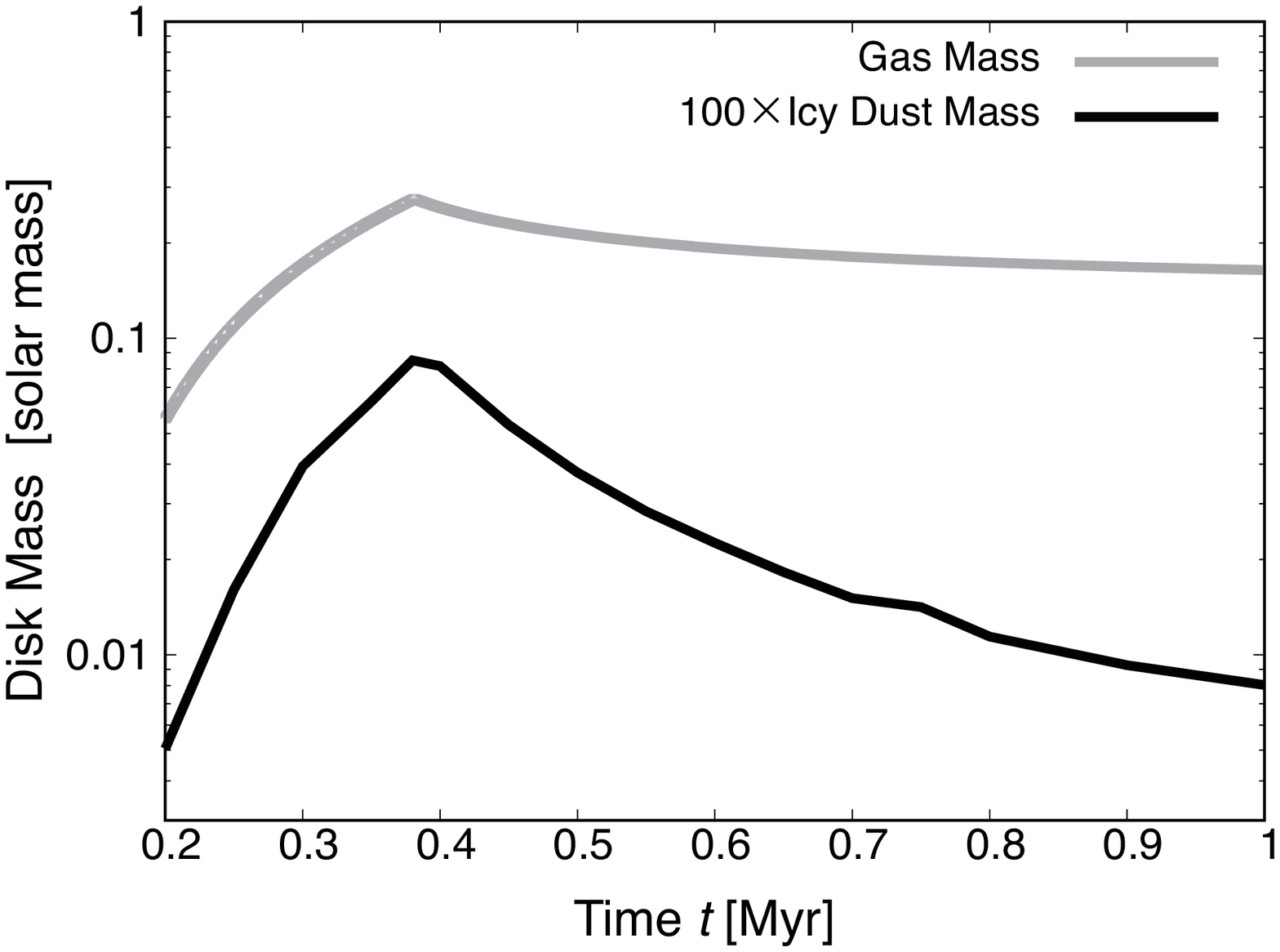}
  \caption{Gas (gray) and icy dust (black) disk masses as a function of the time.
  Note that icy dust disk mass is the mass outside the snowline and multiplied by 100.}
  \label{fig:figdiskevo}
\end{figure}

The radial drift of dust aggregates in the disk formation stage causes the depletion of dust after the infall phase.
Time evolution of the gas and icy dust disk masses is displayed in Figure \ref{fig:figdiskevo}. 
Note that icy dust disk mass is multiplied by 100 to make it easy to compare with the gas disk mass. 
We can see that icy dust depletes more quickly than gas from the end of infall because the drift timescale of dust is smaller than viscous timescale of gas.
This shows that no planetesimal forms after the infall phase if the dust aggregates coagulate and drift in the disk formation stage and dust-to-gas ratio becomes one or more orders of magnitude smaller than solar abundance at 1 Myr.

\subsubsection{Internal Density Evolution}
\begin{figure}[h] 
  \centering
  \includegraphics[width=0.45\textwidth]{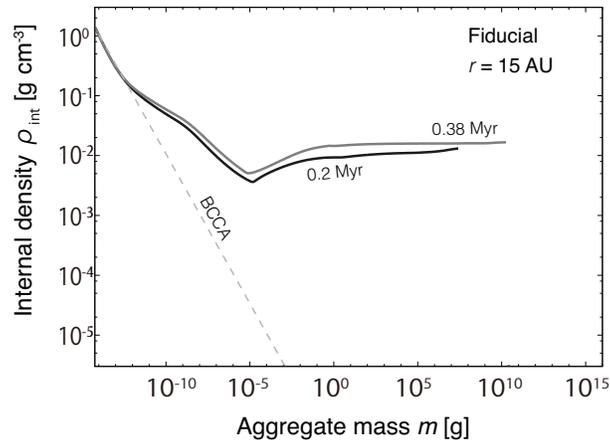}
  \caption{Internal density $\rho_{\rm int} $ at $r=5 {\rm AU}$ at 0.2 and 0.38 Myr for the evolutionary disk fiducial model as a function of the aggregate mass $m$ (solid curves). The dashed line shows the internal density of the BCCA model without compression.}
  \label{fig:fig5}
\end{figure}

Figure \ref{fig:fig5} shows the internal density of the aggregates at $r= 15 \ {\rm AU}$ at $t = 0.2$ and $3.8 \ {\rm Myr}$.
It is seen that the internal density of the aggregates corresponds to that of a BCCA model (dashed line) for very small sized aggregates ($m<10^{-12} \ {\rm g}$).
However, for larger sized aggregates ($10^{-12} \ {\rm g} < m < 10^{-5} \ {\rm g}$),
the internal density is higher than that of the BCCA model
because the contribution to the growth in this size range is dominated by monomer particles.
When the aggregate size reaches $m > 10^{-5} \ {\rm g}$, the internal density is almost 
independent of the aggregate mass because collisional compression becomes effective.

\subsubsection{Projectile Mass Distribution}
\begin{figure}[h] 
  \centering
  \includegraphics[width=0.45\textwidth]{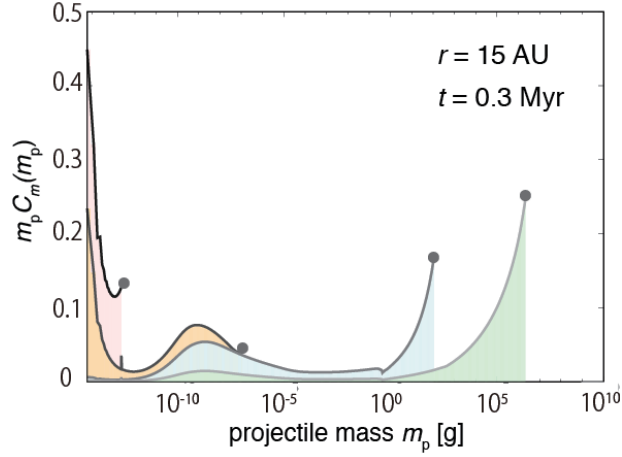}
  \caption{Projectile mass distribution per unit logarithmic projectile mass: $m_{\rm p}C_{m}(m_{\rm p})$ for different targets with mass $m$ at $r=15 \ {\rm AU}$ and $t= 0.3 \ {\rm Myr}$. The circles show the points of equally sized aggregation (i.e., $m_{\rm p} = m$). 
  }
  \label{fig:figprodis}
\end{figure}

Figure \ref{fig:figprodis} shows the projectile mass distribution per unit $\ln m_{\rm p}$ for different targets with mass $m$ at $r=15 \ {\rm AU}$ and $t= 0.3 \ {\rm Myr}$.
When the target mass $m < 10^{-5} \ {\rm g}$, the growth of the target receives contribution from aggregates with a similar mass $m$ and monomer particles with the mass $m_{0}$ (corresponding to the lower mass limit of Figure \ref{fig:figprodis}).
This is because there is a sufficient supply of monomers from the molecular cloud core in this stage.
When the target mass $m > 10^{0} \ {\rm g}$, a variety of aggregates with various masses
contribute to the growth of the target aggregates, but  projectile aggregates with similar mass
are the dominant contributors.

\subsection{Other Models: Effects of the Model Parameters}
To see the physical conditions of the forming and evolving disks for the growth of icy dust aggregates,
the results of other models with different parameters are presented here.
The explored physical conditions include the strength of the turbulence in the disk (turbulence parameter $\alpha_{\rm turb}$);
the initial angular velocity of the molecular cloud core ($\omega_{\rm cd}$), which controls the size of the disk;
and the initial temperature of the molecular cloud core ($T_{\rm cd}$), which alters the mass accretion rate
from the molecular cloud core.
The model parameters examined here are listed in Table \ref{tab:tab1}.

\begin{figure}[h] 
\centering
  \includegraphics[width=0.45\textwidth]{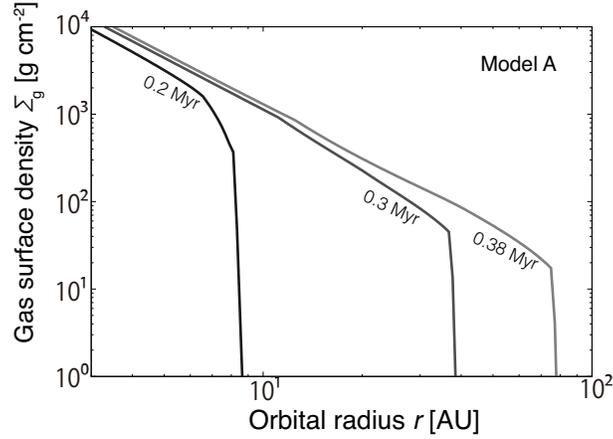}
  \caption{Gas surface density $\Sigma_{\rm g }$ at different times for model A as a function of the orbital radius $r$. }
  \label{fig:figa-gsden}
  \end{figure}

\begin{figure}[h]
  \centering
  \includegraphics[width=0.45\textwidth]{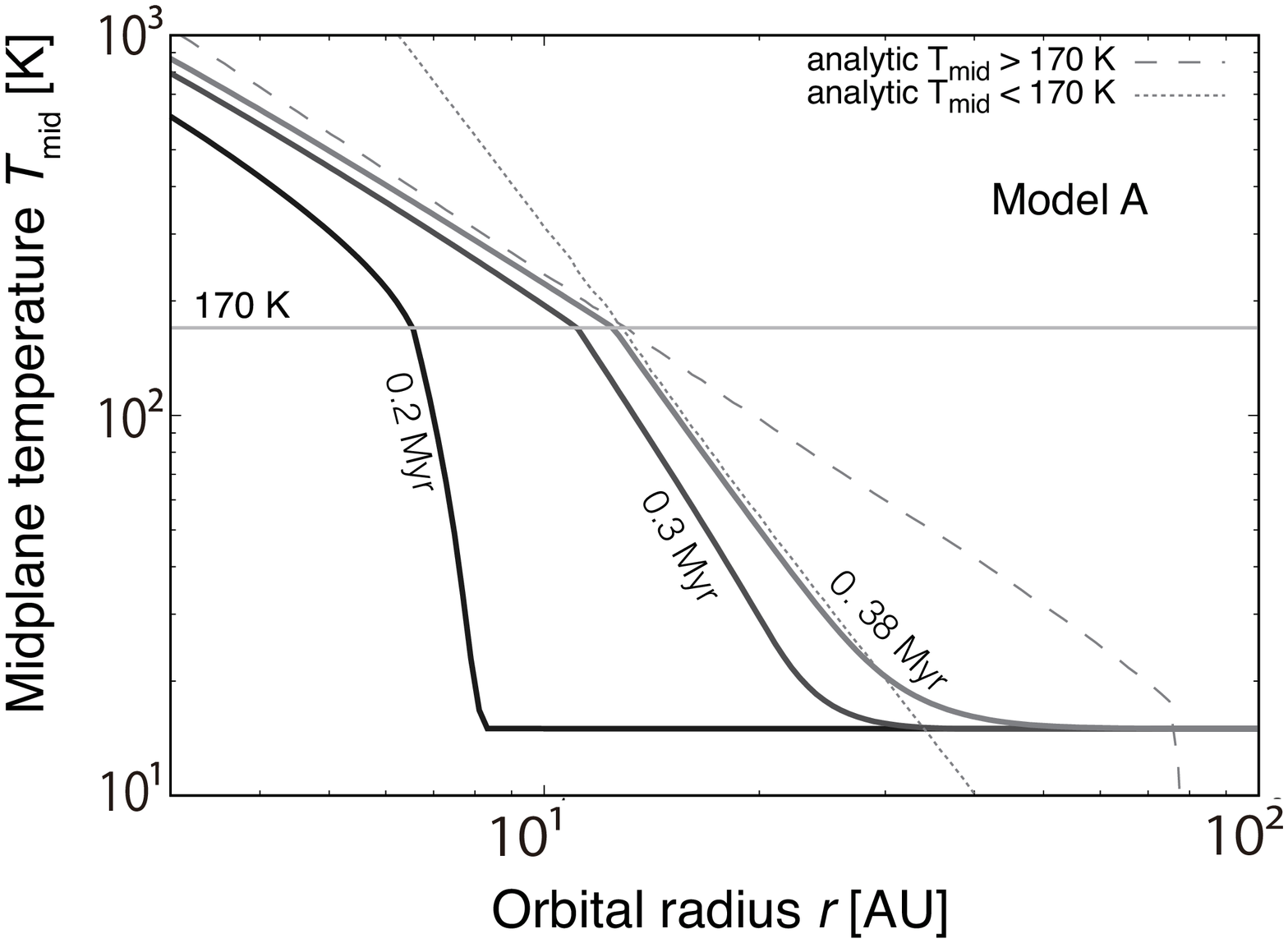}
  \caption{Midplane temperature $T_{\rm mid}$ at each time as a function of the orbital radius $r$ (solid curves). The horizontal dashed line shows $T= 170 \ {\rm K}$. The gray dashed and dotted curves show the analytical solutions of the midplane temperature at 0.38 Myr with ${\dot M}_{\rm star}=10^{-6} \ M_{\odot} \ {\rm yr^{-1}}$.}
  \label{fig:figa-temp}
\end{figure}

\begin{figure}[h]
  \centering
    \includegraphics[width=0.45\textwidth]{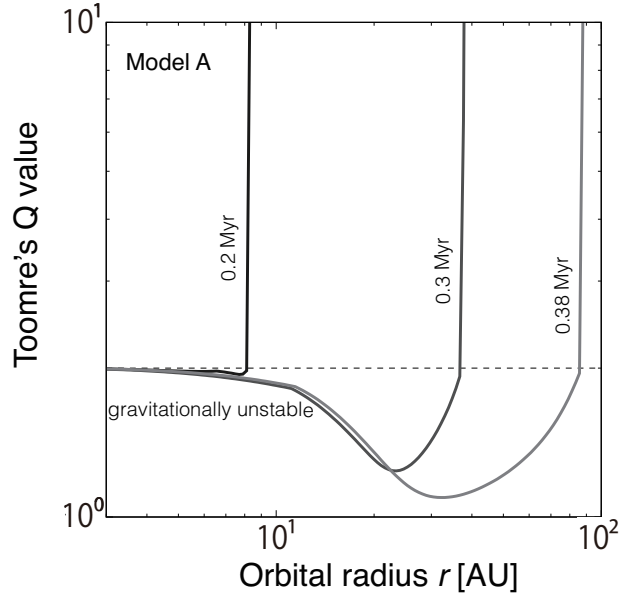}
  \caption{Toomre's ${\cal Q}$ value at different times for model A as a function of the orbital radius $r$ (solid curves). The dashed line shows ${\cal Q}_{\rm cr} = 2 $.}
  \label{fig:figa-toomeq}
\end{figure}

Model A is a weaker turbulence model ($\alpha_{\rm turb} = 10^{-4}$).
The weaker turbulence leads to a lower mass accretion rate of the disk to the central star.
Hence, the gas surface density is likely to be higher than that of a strong turbulence model.
However, when the disk surface density is sufficiently high or the temperature is too low,
the disk becomes gravitationally unstable, and large-scale angular momentum transport occurs. 
The gas surface density $\Sigma_{\rm g }$, the temperature $T_{\rm mid}$ at the midplane, and Toomre's ${\cal Q}$ value at different times for model A are displayed in Figures \ref{fig:figa-gsden}, \ref{fig:figa-temp}, and \ref{fig:figa-toomeq}, respectively.

\begin{figure}[h]
  \centering
    \includegraphics[width=0.45\textwidth]{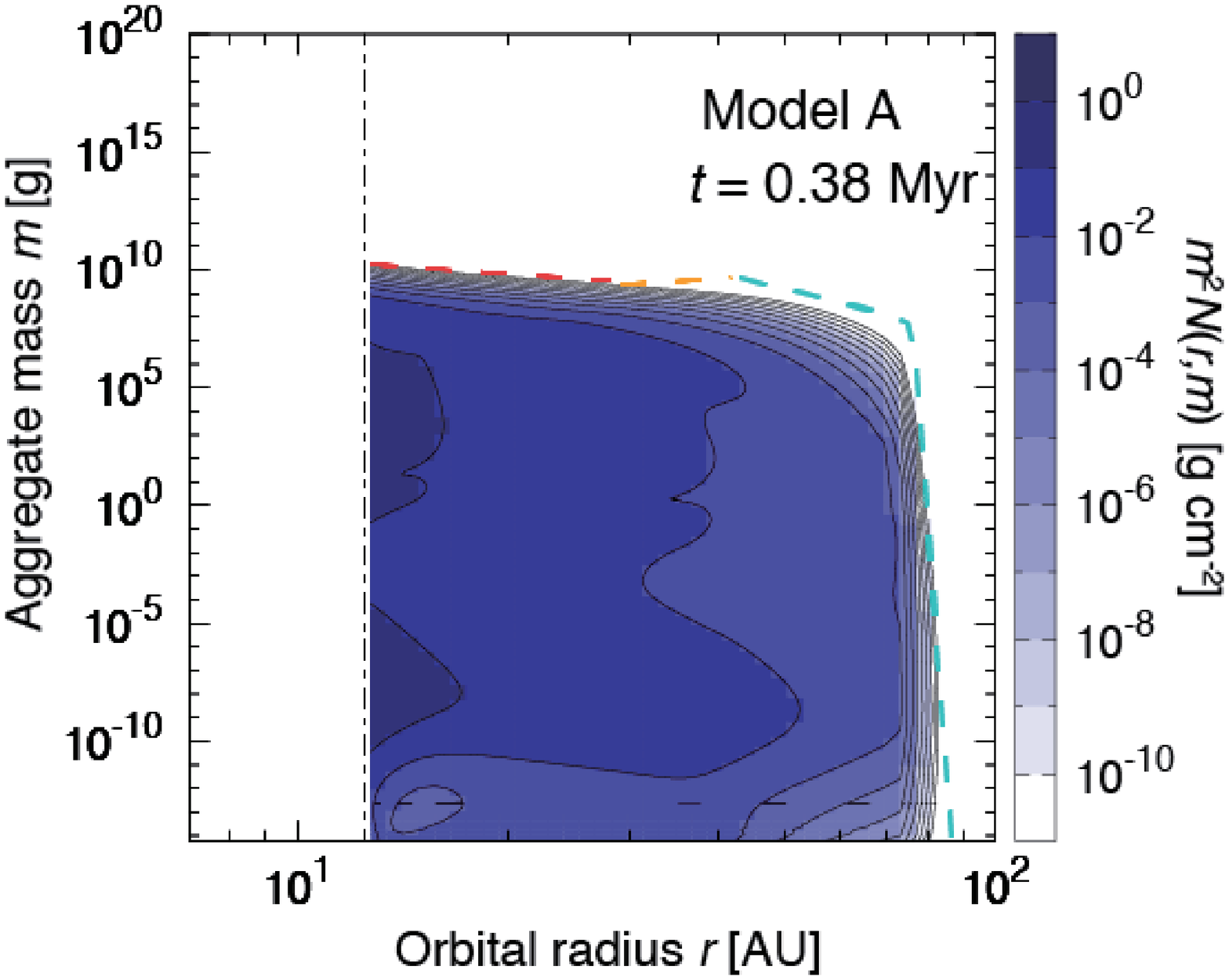}
  \caption{Aggregate size distribution $m^2 {\cal N}$ at different times for model A as a function of the orbital radius (from 7 AU to 100 AU) $r$ and aggregate mass $m$. The dash-dot line shows the snowline.}
  \label{fig:figa-distr}
\end{figure}

In Figure \ref{fig:figa-gsden}, the surface density of the gas is higher than that of the fiducial model
(see Figure \ref{fig:fig3}) in the inner region of the disk ($r<6 \ {\rm AU}$),
but in the outer region ($6 \ {\rm AU} <r$), the surface density is almost the same as that of the fiducial model
because the outer regions of the fiducial model and model A are both gravitationally unstable (see Figures \ref{fig:figQ} and \ref{fig:figa-toomeq}).
This result means that the gas surface density of the fiducial model reaches its maximum in the region where the disk is gravitationally unstable,
and even if the turbulence is weak in the disk formation stage, the gas surface density does not reach the higher value of the fiducial model in the region where icy dust can exist (i.e., $T_{\rm mid}<170 \ {\rm K}$, see Figures \ref{fig:figa-temp} and \ref{fig:figa-distr}).
The growth conditions strongly depend on the dust surface density.
However, as shown above, the gravitationally unstable disk has the maximum gas and dust surface density.

\begin{figure}[h]
  \centering
  \includegraphics[width=0.45\textwidth]{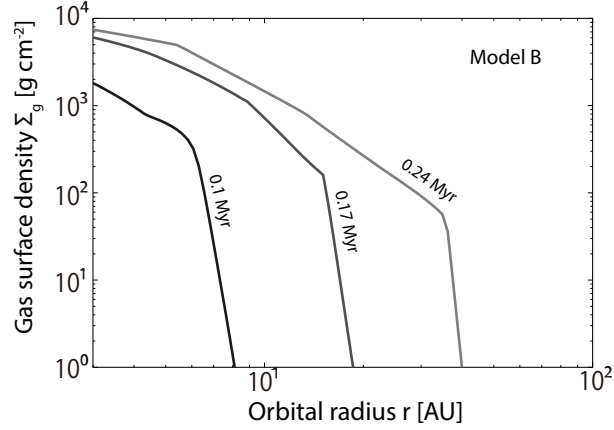}
  \caption{Gas surface density $\Sigma_{\rm g }$ at different times for model B as a function of the orbital radius $r$. }
  \label{fig:figb-gsden}
  \end{figure}

\begin{figure}[h]
  \centering
  \includegraphics[width=0.45\textwidth]{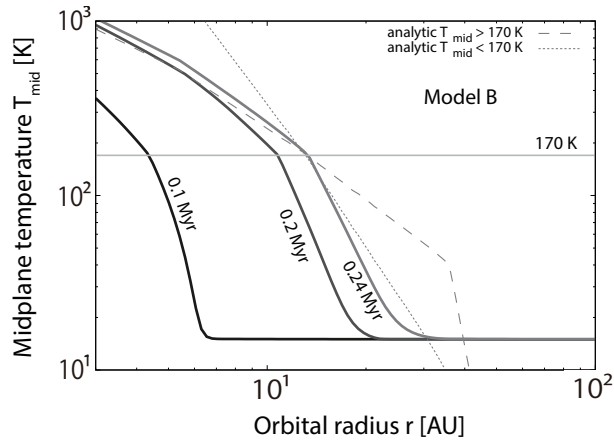}
  \caption{Midplane temperature $T_{\rm mid}$ at each time as a function of the orbital radius $r$ (solid curves). The horizontal dashed line shows $T= 170 \ {\rm K}$. The gray dashed and dotted curves show the analytical solutions of the midplane temperature at 0.24 Myr with ${\dot M}_{\rm star}=10^{-6} \ M_{\odot} \ {\rm yr^{-1}}$.}
  \label{fig:figb-temp}
\end{figure}

In Model B, the temperature of the molecular cloud core is $T_{\rm cd} = 20 \ {\rm K}$, which is higher than that in the fiducial model,
and the higher temperature of the molecular cloud core leads to a higher mass accretion rate from the molecular cloud core.
In this case, the mass accretion lasts about 0.24 Myr, and the centrifugal radius is reduced.
This means that most of the matter from the molecular cloud core falls intensively on a smaller disk.
As a result, the region with accretion from the molecular cloud core is heated more by viscous heating,
the temperature there exceeds 170 K, and all materials fall toward the inside of the snowline (Figures \ref{fig:figb-gsden} and \ref{fig:figb-temp}).

\begin{figure}[h]
  \centering
  \includegraphics[width=0.45\textwidth]{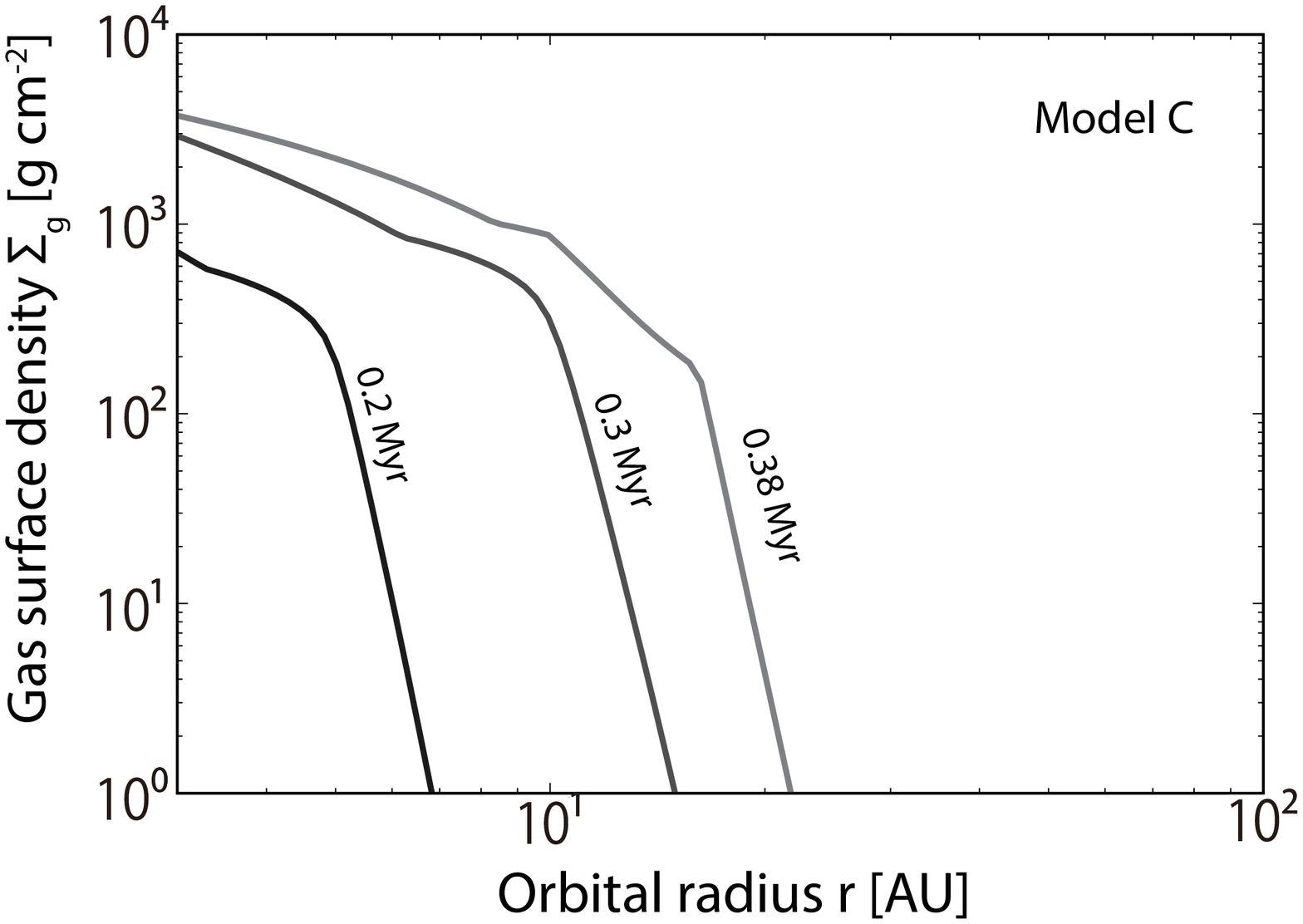}
  \caption{Gas surface density $\Sigma_{\rm g }$ at different times for model C as a function of the orbital radius $r$. }
  \label{fig:figc-gsden}
  \end{figure}

\begin{figure}[h]
  \centering
  \includegraphics[width=0.45\textwidth]{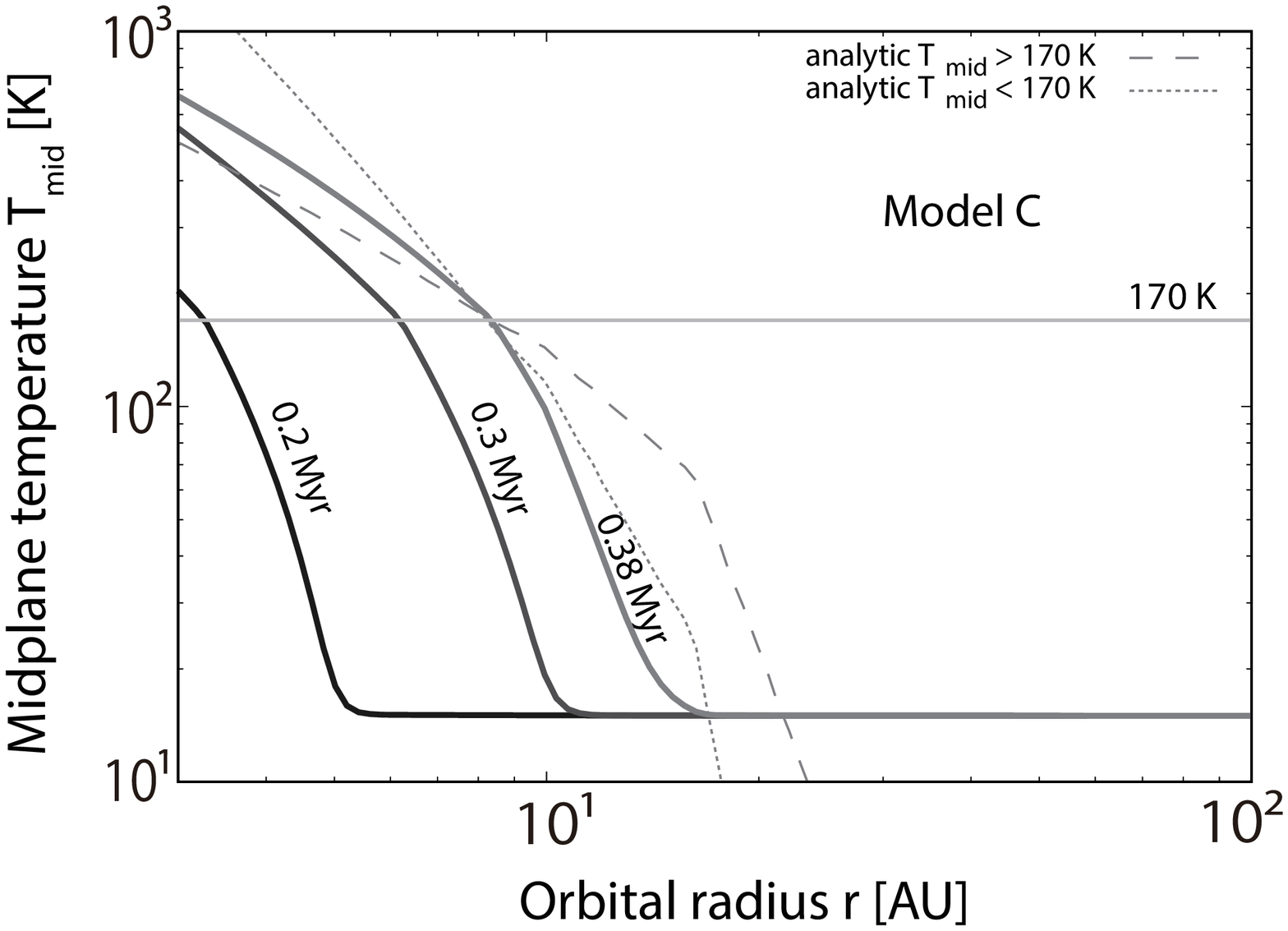}
  \caption{Midplane temperature $T_{\rm mid}$ at each time as a function of the orbital radius $r$ (solid curves). The horizontal dashed line shows $T= 170 \ {\rm K}$. The gray dashed and dotted curves show the analytical solutions of the midplane temperature at 0.38 Myr with ${\dot M}_{\rm star}=10^{-6} \ M_{\odot} \ {\rm yr^{-1}}$.}
  \label{fig:figc-temp}
\end{figure}

The results for Model C are similar to those for Model B.
In Model C, the initial angular velocity of the molecular cloud core is $\omega_{\rm cd}=5 \times 10^{-15} \ {\rm s^{-1}}$, which is slower than that of the fiducial model.
The lower angular velocity leads to a smaller centrifugal radius, and as a result, most of the matter falls to the smaller disk (Figures \ref{fig:figc-gsden} and \ref{fig:figc-temp}). 

\begin{figure}[h]
  \centering
    \includegraphics[width=0.45\textwidth]{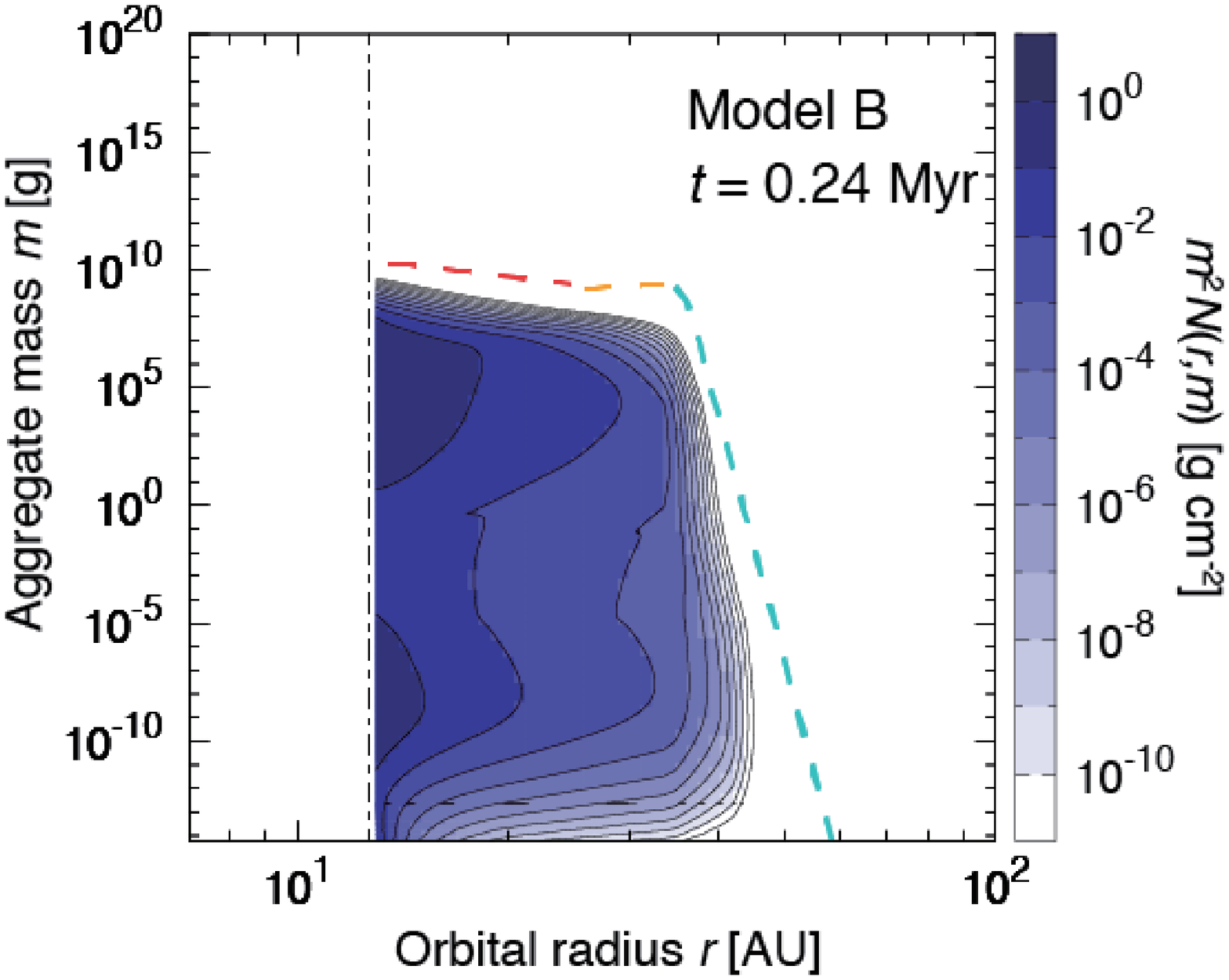}
  \caption{Aggregate size distribution $m^2 {\cal N}$ at different times for Model B as a function of the orbital radius (from 7 AU to 100 AU) $r$ and aggregate mass $m$. }
  \label{fig:figc-distr}
\end{figure}

\begin{figure}[h]
  \centering
    \includegraphics[width=0.45\textwidth]{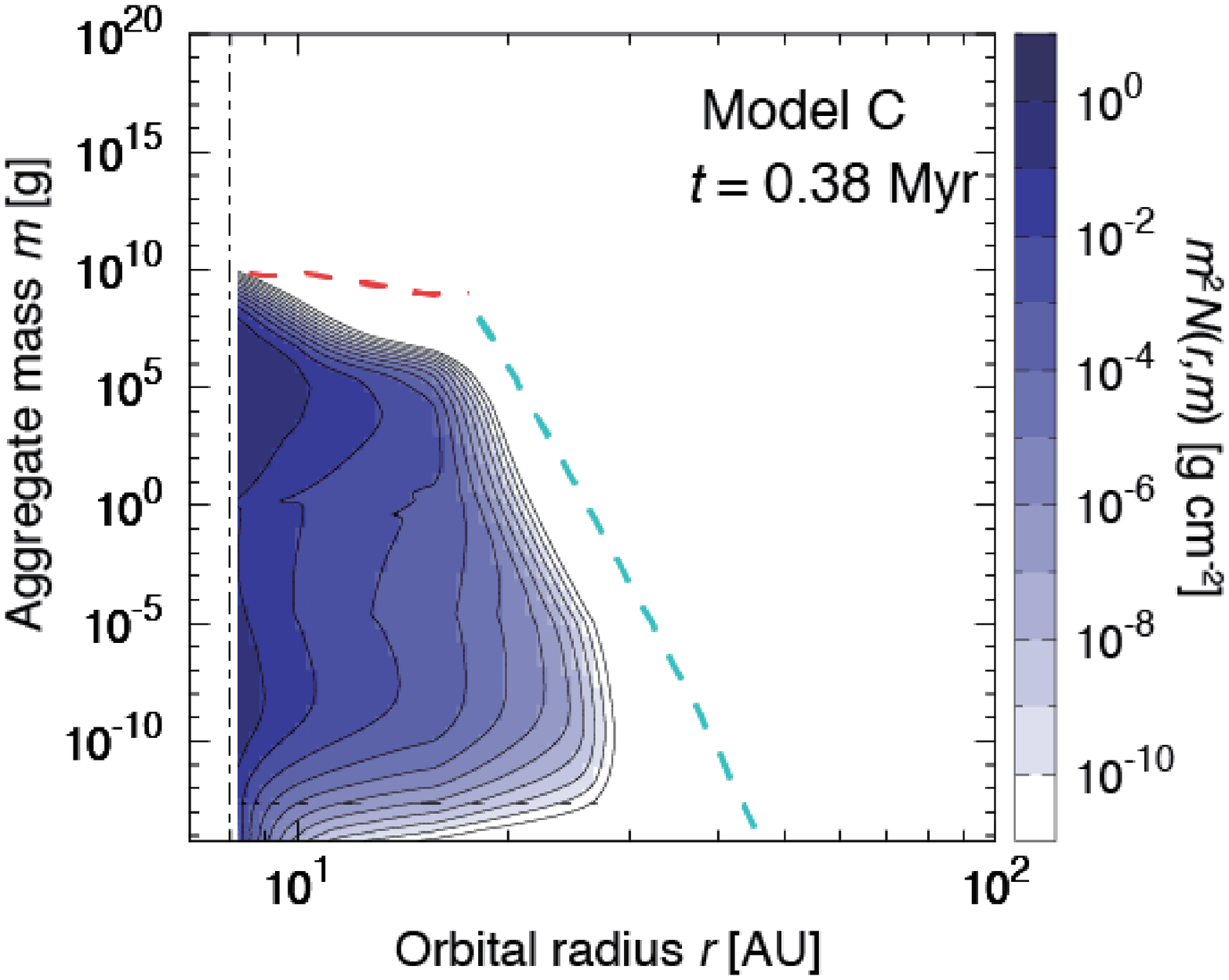}
  \caption{Aggregate size distribution $m^2 {\cal N}$ at different times for model C as a function of the orbital radius (from 7 AU to 100 AU) $r$ and aggregate mass $m$. }
  \label{fig:figd-distr}
\end{figure}

Figures \ref{fig:figc-distr} and \ref{fig:figd-distr} show the size distribution of the aggregates for Models B and C at the end of the mass accretion from the molecular core.
In these cases, there are no aggregates exceeding the size of $\Omega t_{\rm ts} =1$. 

The results with these different parameters (Models A, B, and C) indicate
that it is difficult for icy dust aggregates to grow to the size of the planetesimal via direct coagulation in a range of reasonable parameters in the disk formation stage.

\newpage
%
\section{DISCUSSION}
\label{sec:discussion}


%
%
\subsection{Difficulties in Breaking Through the Radial Drift Barrier in the Disk Formation Stage}
\label{sec:secgcon}
Here, comparing the timescales of aggregate growth and radial drift,
we explore the reason why icy aggregates cannot break through the radial drift barrier in the disk formation stage.
When growth occurs mainly through collisions with similarly sized aggregates,
the growth rate of an aggregate with mass $m$ at the midplane is given by
$\frac{dm}{dt} = \frac{\Sigma_{\rm d}}{\sqrt{2 \pi} h_{\rm d}} \sigma_{\rm coll} \Delta v$. 
Then, the timescale of the aggregate growth is written as
$t_{\rm grow} \equiv \frac{m}{dm/dt} = 
\sqrt{2 \pi} \frac{h_{\rm d}}{\Delta v} \frac{m/ \sigma_{\rm coll}}{\Sigma_{\rm d}}$. 
Using $m=(4 \pi/3) \rho_{\rm int} a^3$ and $\sigma_{\rm coll} = \pi a^2$, we have
\begin{equation}
t_{\rm grow} = \frac{4 \sqrt{2 \pi}}{3} \frac{h_{\rm d}}{\Delta v} \frac{\rho_{\rm int} a}{\Sigma_{\rm d}}.
\end{equation}

Now, we focus on aggregates of the size corresponding to $\Omega t_{\rm s}=1$
because the radial drift velocity reaches the maximum value at this size.
For dust aggregates with this size,
the scale height of the dust disk is given by $h_{d\rm } \approx \sqrt{\alpha_{\rm turb}}h_{\rm g} $ according to Eq. (\ref{eq:eq10}),
and we set the relative velocity $\Delta v \approx \sqrt{\alpha_{\rm turb}} c_{\rm s}$
because the collisional velocity is dominated by the turbulence-driven velocity described by Eq. (\ref{eq:eqdelvt}) at $\Omega t_{\rm s}=1$.
Then, we can write the growth timescale at $\Omega t_{\rm s} =1$ as
\begin{equation}
t_{\rm grow}|_{\Omega t_{\rm s}=1}
= \frac{4 \sqrt{2 \pi}}{3}\frac{(\rho_{\rm int} a)_{\Omega t_{\rm s}=1}}{\Sigma_{\rm d}\Omega}
= \frac{4 }{3 \sqrt{2 \pi}}\frac{(\rho_{\rm int} a)_{\Omega t_{\rm s}=1}}{\Sigma_{\rm d}} t_{\rm K},
\end{equation}  
where $h_{\rm g} = c_{\rm s}/\Omega$ and the Kepler orbital period $t_{\rm K} = 2 \pi /\Omega$ are used.
When the dust aggregates are influenced by the Stokes law ($Re_{\rm p} < 1$),
the timescale for the aggregate at $\Omega t_{\rm s} =1$ is given by
\begin{eqnarray}
\label{eq:eqt1}
t_{\rm grow}|_{\Omega t_{\rm s}=1}^{({\rm St})} &=& 1.2 \times 10^2
\biggl( \frac{\rho_{\rm int}}{10^{-2} \ {\rm g \ cm^{-3}}}\biggl)^{\frac{1}{2}}  
\biggl( \frac{M}{M_{\odot}}  \biggr)^{-\frac{3}{4}} \\ \nonumber
 &&\times
\biggl( \frac{ T_{\rm mid}}{170 \ {\rm K}}  \biggr)^{\frac{1}{4}} 
\biggl( \frac{r}{15 {\rm AU}}  \biggr)^{\frac{9}{4}} 
\biggl( \frac{\Sigma_{\rm d}/{\Sigma_{\rm g}}}{0.01}\biggr)^{-1} 
\biggl( \frac{\Sigma_{\rm g}}{10^{3} \ {\rm g \ cm^{-2}}}\biggr)^{-1}  \ {\rm yr} , 
\end{eqnarray} 
and when the aggregates are controlled by the Allen law ($1 < Re_{\rm p} < 800$),
it is given by
\begin{eqnarray}
t_{\rm grow}|_{\Omega t_{\rm s}=1}^{({\rm Al})} &=& 1.8 \times 10^2
\biggl( \frac{\rho_{\rm int}}{10^{-2} \ {\rm g \ cm^{-3}}}\biggl)^{\frac{3}{8}} 
\biggl(\frac{\eta}{5\times10^{-3}} \biggr)^{\frac{1}{4}}  \nonumber \\ 
                        &&\times
                    \biggl( \frac{M}{M_{\odot}}  \biggr)^{-\frac{9}{16}} \biggl( \frac{ T_{\rm mid}}{170 \ {\rm K}}  \biggr)^{\frac{1}{16}} 
                    \biggl( \frac{r}{15 {\rm AU}}  \biggr)^{\frac{31}{16}} \biggl( \frac{\Sigma_{\rm d}/{\Sigma_{\rm g}}}{0.01}\biggr)^{-1} 
                    \biggl( \frac{\Sigma_{\rm g}}{10^{3} \ {\rm g \ cm^{-2}}}\biggr)^{-\frac{3}{4}}  \ {\rm yr}.
\end{eqnarray}
On the other hand, the radial drift timescale at $\Omega t_{\rm s}=1$ is given by
\begin{eqnarray}
\label{eq:eqtdr}
t_{\rm drift}|_{\Omega t_{\rm s}=1} &=& \frac{r}{|v_r|_{\Omega t_{\rm s}=1}} =\frac{1}{2 \pi \eta} t_{\rm K} \nonumber \\ 
&=&1.9 \times 10^3 \biggl(\frac{\eta}{5\times10^{-3}} \biggr)^{-1}  \biggl( \frac{M}{M_{\odot}}  \biggr)^{-\frac{1}{2}} \biggl( \frac{r}{15 {\rm AU}}  \biggr)^{\frac{3}{2}} \ {\rm yr}, 
\end{eqnarray} 
where $|v_r|_{\Omega t_{\rm s}=1} = \eta v_{\rm K}$ is used.

\begin{figure}[h] 
  \centering
    \includegraphics[width=0.45\textwidth]{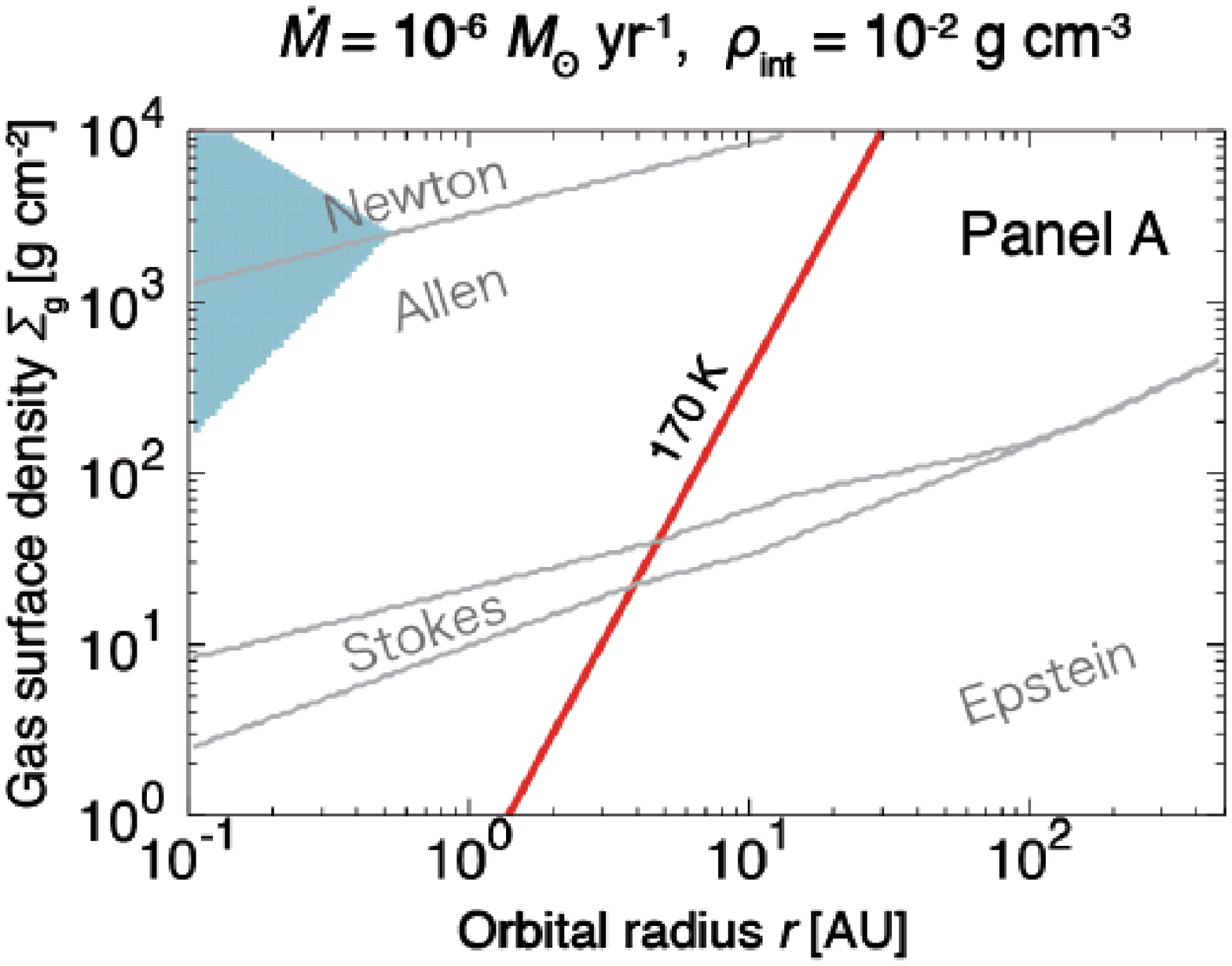}
      \includegraphics[width=0.45\textwidth]{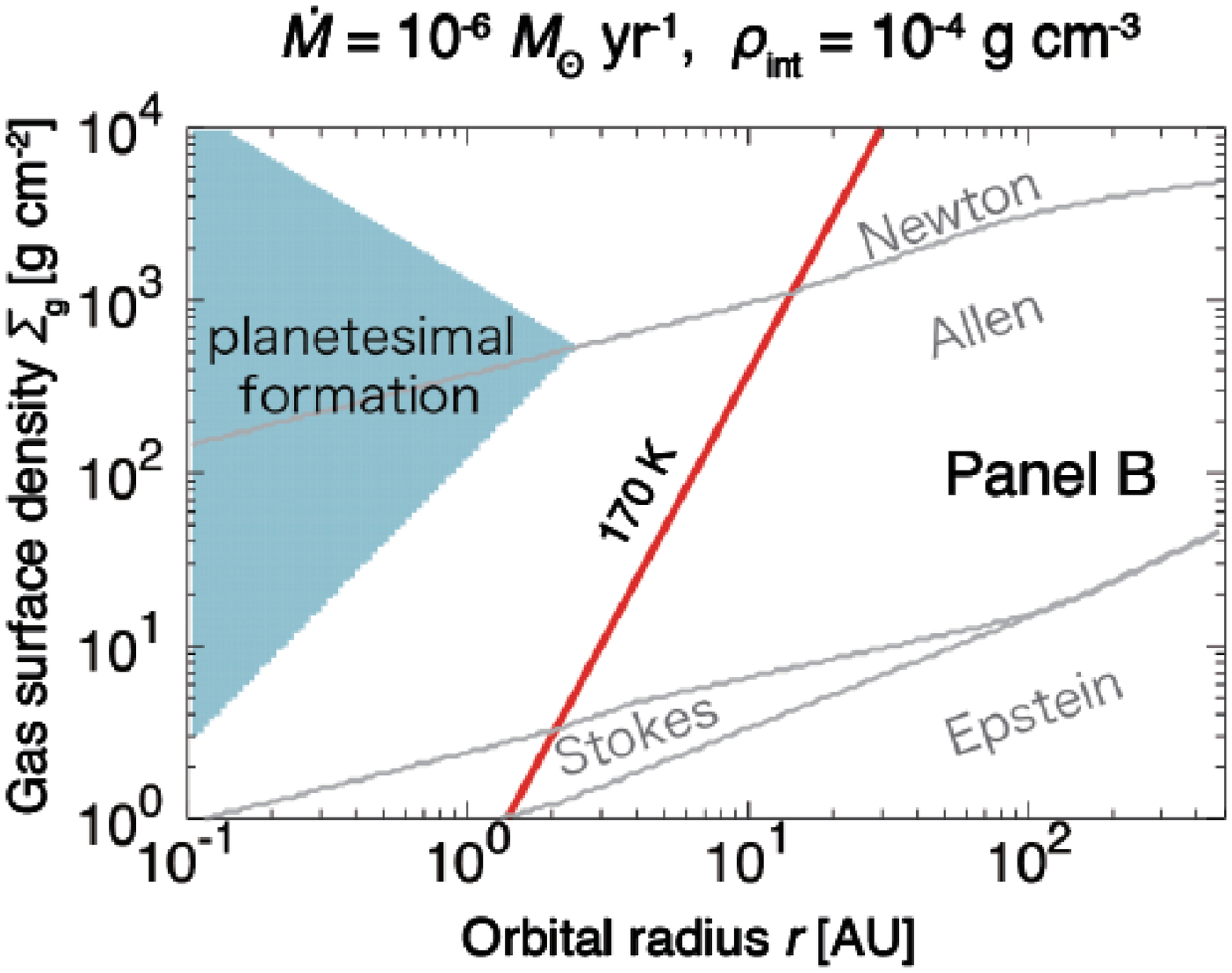}
      
  \caption{Growth condition diagram at $\Omega t_{\rm s}=1$ with an accretion rate of $10^{-6} \ M_{\odot} \ {\rm yr^{-1}}$ and an internal density of dust aggregates $\rho_{\rm int} = 10^{-2} \ {\rm g \ cm^{-3}}$ (panel A) and  $\rho_{\rm int} = 10^{-4} \ {\rm g \ cm^{-3}}$ (panel B). The blue region shows cases where $t_{\rm grow}/t_{\rm drift}<1/30$, and planetesimal formation is expected to take place. The horizontal axis is the orbital radius, and the vertical axis is the gas surface density. The red solid line shows $T=$170 K, and the gray solid curves show the boundaries among the gas drag laws.}
  \label{fig:growthcon}
\end{figure}

We adopt the growth condition given by \cite{okuzumi2012}: 
\begin{equation}
\label{eq:eqgcon}
\biggl( \frac{t_{\rm grow}}{t_{\rm drift}} \biggr)_{\Omega t_{\rm s}=1} < \frac{1}{30},  
\end{equation}
which was derived from the results of numerical calculations. 
Now, we consider the steady accretion disk because the results show an almost constant accretion rate  ${\dot M} \sim 10^{-6} \ M_{\odot} \ {\rm yr^{-1}}$ in the region on which we focus.
Considering an optically thick disk with steady accretion and a dust-to-gas mass ratio $\Sigma_{\rm d}/{\Sigma_{\rm g}}=0.01$, the analytical solutions of the temperature of the disk midplane are given by Eq. (\ref{eq:anatemp}) using Eqs. (\ref{eq:midtemp}) and (\ref{eq:opacity}).
If we use the appropriate ${\dot M}$, this analytical solution is a good result of the numerical calculation.
Using Eqs. (\ref {eq:anatemp}) and (\ref{eq:eqgcon}), we investigate the region where dust can increasingly grow to the size of a planetesimal, breaking through the radial drift barrier in the $ r - \Sigma_{\rm g}$ space, and show it in Figure \ref{fig:growthcon}, where we use ${\dot M} \sim 10^{-6} \ M_{\odot} \ {\rm yr^{-1}}$, $\eta=2\times c_{\rm s}^2/v_{\rm K}^2$, $\Sigma_{\rm d}/{\Sigma_{\rm g}}=0.01$, and $M = 1M_{\odot}$.
The panel A shows the case of $\rho_{\rm int} = 10^{-2} \ {\rm g \ cm^{-3}}$ based on our numerical calculation results. 
We can see that the region where the dust aggregates can grow by breaking through the radial drift barrier at the size of $\Omega t_{\rm s}=1$ (blue colored region) only exists inside the snowline.
This result indicates that there is no icy dust that can grow to planetesimals beyond the radial drift barrier via direct collisional growth in the disk, where mass accretion to the central star is large, and viscous heating is sufficiently effective.

The panel B of Figure \ref{fig:growthcon} shows the case of $\rho_{\rm int} = 10^{-4} \ {\rm g \ cm^{-3}}$, which is the internal density achieved at $\Omega_{t_{\rm s}}=1$ with aggregates composed of 0.1-${\rm \mu m}$ icy monomers when static gas compression is effective \citep{kataoka2013b}. 
Even in this case, as in the case of $\rho_{\rm int} = 10^{-2} \ {\rm g \ cm^{-3}}$, the area where icy dust can grow to planetesimals is limited inside the snow line. 
Although our model of internal density evolution by collisional compression is very simple, as can be seen from these analytical estimates, the internal density of the aggregates does not affect our calculation results.

\subsection{Comparison with Tsukamoto et al. (2017)}
The previous section showed the difficulty of ice planet formation at the disc formation stage. 
On the other hand, \cite{tsukamoto2017}, which can be cited as a similar study to our study, assumed that the gas disk is gravitationally unstable in the disk formation stage and investigated the collisional growth of icy dust in a gravitationally unstable steady accretion disk.
One of their results is that the maximum  orbital radius within which icy planetesimals form via the coagulation of porous icy dust aggregates becomes $r \sim 20$ AU in the gravitationally unstable disk around a solar mass star.

As we have seen from the previous section, our results show that it is difficult to form icy planetesimals, although our model using disk evolution including mass accretion from a molecular cloud core also shows a gravitationally unstable disk. 
This difference comes from the difference in the models of the gas friction law of an aggregate with a high Reynolds number.
When the particle Reynolds number becomes larger than unity, the aggregate suffers gas friction, which is called Allen's law, and when it becomes nearly $10^3$, the gas friction follows Newton's law.
However, these friction laws are not considered in \cite{tsukamoto2017}.

\begin{figure}[h]
  \centering
    \includegraphics[width=0.45\textwidth]{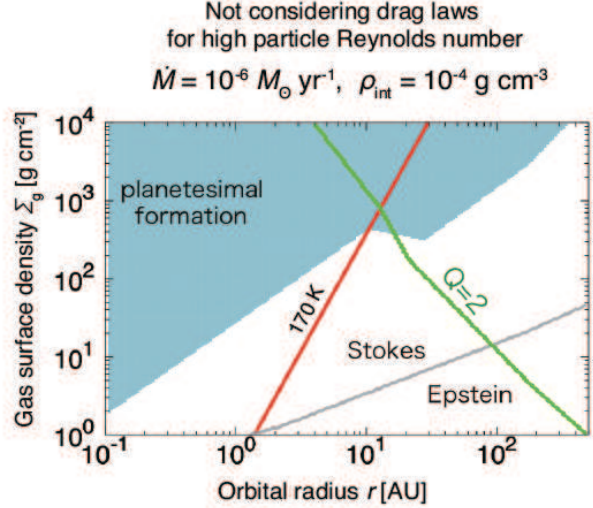}
  \caption{Growth condition diagram at $\Omega t_{\rm s}=1$ with an accretion rate of $10^{-6} \ M_{\odot} \ {\rm yr^{-1}}$ and an internal density of dust aggregates $\rho_{\rm int} = 10^{-4} \ {\rm g \ cm^{-2}}$. The blue region shows $t_{\rm grow}/t_{\rm drift}<1/30$. Note that we do not consider the gas drag laws for a high Reynolds number. The horizontal axis is the orbital radius, and the vertical axis is the gas surface density. The red solid line shows $T=$170 K, the gray solid line shows the boundary between Stokes law and Epstein's law, and the green solid curve shows the gas surface density with $Q=2$. }
  \label{fig:growthcon2}
\end{figure}

Figure \ref{fig:growthcon2} is a diagram of the growth conditions under the same conditions the panel B of Figure \ref{fig:growthcon}, except that it only considers the gas friction law as Stokes law.
As we can see in Figure \ref{fig:growthcon2}, if we do not consider Allen's law and Newton's law, the region that can avoid the radial drift barrier (blue colored region) extends to the outer region of the snow line.
The green curve in Figure \ref{fig:growthcon2} is the gas surface density when $Q = 2$, which indicates a gravitationally unstable disk.
Comparing the blue region and green line in Figure \ref{fig:growthcon2}, it is found that the maximum orbital radius of the formation of planetesimals reaches approximately 20 AU, and this estimate is consistent with the result of \cite{tsukamoto2017}. 

Comparing our results and those obtained by \cite{tsukamoto2017}, we can find that it is very important to consider the friction law for a high particle Reynolds number when we consider a heavy disk and highly porous aggregates since the particle Reynolds number is proportional to the radius of the aggregate and the gas density.

%
%
\subsection{Possibilities for Icy Planetesimal Formation}
\label{sec:possibility}
Our results show that the icy dust aggregates suffer radial drift to the central star when they reach $\Omega t_{\rm s}\sim 1$ in the disk formation stage. 
Once the radial drift of aggregates is dominant, the dust supply and radial drift are balanced and no icy planetesimal is formed. 
In this section, we discuss the possibilities for icy planetesimal formation. 
One possibility is the process of keeping the icy dust small without growing to a large size during the disk formation stage. 
A second possibility is  that the temperature becomes lower than the model we are considering owing to the decrease in the opacity with the dust size evolution. 
The last one is that planetesimals are formed by processes other than direct coagulation, such as the streaming instability.
We will discuss these possibilities in the following.

\subsubsection{Processes Keeping the Dust Small in a Protoplanetary Disk}
\begin{figure}[h]
  \centering
    \includegraphics[width=0.45\textwidth]{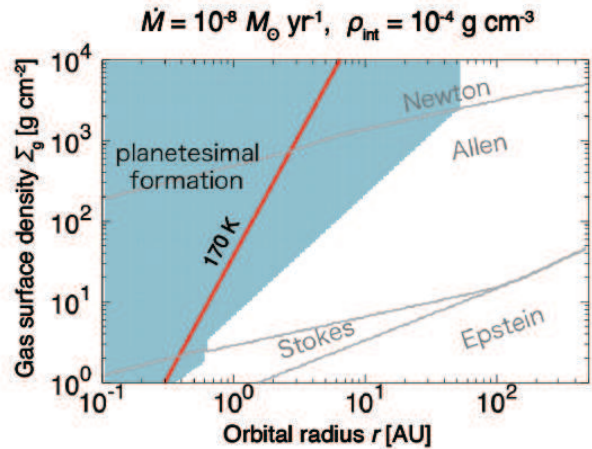}
  \caption{Growth condition diagram at $\Omega t_{\rm s}=1$ with an accretion rate of $10^{-8} \ M_{\odot} \ {\rm yr^{-1}}$ and an internal density of dust aggregates $\rho_{\rm int} = 10^{-4} \ {\rm g \ cm^{-3}}$. The blue region shows $t_{\rm grow}/t_{\rm drift}<1/30$. The horizontal axis is the orbital radius, and the vertical axis is the gas surface density. The red solid line shows $T=$170 K, and the gray solid curves show the boundaries among the gas drag laws.}
  \label{fig:growthcon3}
\end{figure}

As we saw in the previous section, the reason why it is difficult for icy dust to grow into planetesimals is that the snow line reaches about 10 AU by viscous heating, and the area where dust can grow beyond the radial drift barrier is limited inside the snow line.
Figure \ref{fig:growthcon3} shows the growth condition diagram when the mass accretion rate toward the central star is $10^{-9} \ M_{\odot} \ {\rm yr^{-1}}$ and the internal density of the aggregate is $\rho_{\rm int} = 10^{-4} \ {\rm g \ cm^{-3}}$.
In this cace, the snow line is located closer to the central star, and the area where dust can grow into planetesimals extends outside the snow line. 
This means that it is necessary to keep the dust close to the monomer size during the disk formation stage in order to form planetesimals by direct coagulation. 

One powerful process that inhibits dust particles from growing to millimeter size is the charge barrier \citep{okuzumi2009a}. 
The charge barrier is a process in which dust particles are negatively charged by capturing the electrons of an ionized gas, and collisional growth for small dust particles is suppressed. \cite{okuzumi2011a} found the region where the negative charging stalls dust collisional growth at the fractal growth stage of coagulation, which is called the ``frozen'' zone. 
\cite{okuzumi2011b} also showed that the global transport of macroscopic aggregates from outside the frozen zone can lead to the removal of frozen aggregates and estimated that the removal timescale reaches $10^6$ yr. 
This time scale is longer than that of disk formation ($\sim10^5$ yr); hence, the charge barrier can be a process that keeps the dust particles very small during the disk formation stage.

Collisional fragmentation is also a process that inhibits dust growth.
Collisional fragmentation is generally recognized as a serious barrier to planetesimal formation.
Assuming icy monomers of $ 0.1 \ {\rm \mu m} $, $ N $-body numerical experiments suggest that the aggregate suffer from catastrophic fragmentation at the collision velocity $\Delta v = 35 - 70 \ {\rm m \ s^{- 1}} $\citep{wada2009}. 
On the other hand, the relative collisional velocity $v_{\rm coll}$ driven by turbulence reaches the maximum when the dust aggregates have $\Omega t_{\rm s}\simeq 1$, and it is roughly given by $v_{\rm coll} \sim \sqrt{\alpha_{\rm turb}} \times c_{\rm s}.$ 
Then the maximum collisional velocity induced by turbulence in the region outside the snowline is about $ 24 \ {\rm m \ s^{-1}}$ with $T_{\rm mid} = 170 {\rm \ K}$ and $\alpha_{\rm turb} = 10^{-3}$.
This implies that the catastrophic fragmentation can be ignored in this model.

Erosion, a process in which the target loses mass through many high-velocity collisions with small projectiles, is another process that hinders the dust growth.
The radial drift velocity depends on the size of dust aggregates,
and the relative radial velocity between dust aggregates having
$\Omega t_{\rm s} = 1$ and much smaller ones is given by $\eta v_{\rm K}$.
The value of $\eta$ in our model is about $10^{-2}$ outside the snowline,
so the relative velocity can be up to 77 m s$^{-1}$ at 15 AU.
It is suggested, however, that the critical velocity for erosive mass loss is 100 m s$^{-1}$ or higher when the monomer is ice and its size is 0.1 $\mu$m \citep{gundlach2015}.
Thus, the mass loss due to the erosion can be ignored in our model as well.

For the mass loss processes, such as the fragmentation and the erosion,
the monomer size of dust aggregates plays an important role.
The sticking efficiency between monomers depends on the monomer size:
the larger the monomer size is, the lower the sticking efficiency becomes.
When the size of icy monomer is 10 $\mu$m,
the critical velocities for fragmentation and for erosion become as low as
a few m s$^{-1}$ \citep{wada2013, gundlach2015}.
Such large monomers, which may be formed by condensation of water vapor at around the snowline \citep{johansen2009}, may lead to an effective destruction of dust aggregates and the depletion of dust due to the radial drift.
These effects should be examined in the future work.

\subsubsection{Influence of the Opacity}
In our model, in order to simplify the calculations, we used the properties of interstellar dust and fixed the dust-to-gas ratio to determine the opacity. 
However, in reality, the size of dust particles changes owing to collisional growth, and the opacity may differ from that of the interstellar one. 
From the results of our calculations, icy dust grows quickly to a macroscopic size; thus our calculations may overestimate the opacity in a protoplanetary disk. 
As discussed in the previous section, the reason why icy dust drifts to the central star is that the disk is warmed by viscous heating, the snow line spreads outward, and the area where dust can grow to a planetesimal size is limited inside the snow line. 
Therefore, the effect of decreasing the opacity due to the growth in the size of the dust may work favorably for planetesimal formation by direct collisional growth.
However, the mass opacity of dust aggregates also depends on their porosity and it can be characterized by $a \times f$, where $f$ is the filling factor and $a$ is the radius of the dust aggregates \citep{kataoka2014}. 
We need to calculate the dust size distribution, porosity, and opacity simultaneously to determine the midplane temperature, which is a future work.  

\subsubsection{Streaming Instability}
A streaming instability is caused by a two-fluid instability due to the relative drift between the dust and the gas in the protoplanetary disk \citep{youdin2005}.
As a result of the streaming instability, dust particles that have sizes close to $\Omega t_{\rm s} \sim 1$
(or even smaller values of $\Omega t_{\rm s}$ are suggested by \cite{carrera2015} and \cite{yang2017})
form dust clumps, and if the density of the dust clump becomes larger than the Roche density, planetesimals are formed by gravitational collapse of those dust clumps \citep[e.g.,][]{johansen2007, johansen2007b}. 
The formation of dust clumps by the streaming instability requires an increase in $\Sigma_{\rm d}/\Sigma_{\rm g}$ by a few times from the solar value of $10^{-2}$ \citep{johansen2009, carrera2015}. 
Processes for achieving an enhancement in the dust-to-gas mass ratio and planetesimal formation by the streaming instability are subjective; the dissipation of the gas from a disk by photoevapration can produce a massive planetesimal belt beyond 100 AU \citep{carrera2017}, and the pile up of drifting macroscopic dust by the dust-to-gas back reaction creates a narrow planetesimal formation region at the inner edge of the protoplanetary disk \citep{drazkowska2016}. 
However, these mechanisms cannot form sufficiently early and sufficient icy planetesimals in the Saturn and Jupiter region (5--10 AU). 
The evaporation and recondensation of water outside the snowline can be the strongest process that triggers an enhancement in the dust-to-gas mass ratio for icy dust in the disk formation stage.
Drifting through the snowline, icy dust evaporates, and the vapor recondenses when it is transferred outside the snowline owing to diffusion.
This process can trigger the streaming instability near the outside of the snowline \citep{schoonenberg2017}. 
\cite{drazkowska2018} investigated planetesimal formation by the streaming instability beginning in the disk build-up phase.
They found that icy planetesimals are formed just outside the snowline due to the re-condensation of water vapor, the radial drift of dust particles/aggregates, and the traffic jam of dust at just after the disk build-up phase. 
The enhancement of dust-to-gas mass ratio just outside the snowline due to those effects is also found in our results.
However, the enhancement seen in our results is not enough to lead to the streaming instability.
This difference seems to be originated from the difference of models used in this study and in \cite{drazkowska2018}:
in \cite{drazkowska2018}, the critical collision velocity for destruction of dust aggregates was assumed to be 10 m s$^{-1}$ and the back reaction from dust to gas is taken into account,
while in the model of this study, the critical velocity is assumed to be so high that
the fragmentation does not happen
and the back reaction is not taken into consideration.
Judging from results by this study by \cite{drazkowska2018},
icy planetesimal formation caused by the streaming instability near the snowline seems to be a promising formation mechanism.
A more precise model including the streaming instability for planetesimal formation in the disk formation stage is desirable in the future work.

\section{SUMMARY}
\label{sec:sum}
We have investigated how disk evolution in the disk formation stage affects the collisional growth and radial motion of porous icy dust aggregates. We have calculated the evolution of the radial size distribution of icy dust aggregates using the disk model developed by \cite{nakamoto1994} and \cite{hueso2005}. Our study is summarized as follows.

\begin{enumerate}
\item The disk temperature rises easily by viscous heating,  the snow line reaches as much as 10 AU at the maximum, and the disk becomes gravitationally unstable in the outer region where the disk is cold and massive in the disk formation stage.
\item For any parameters related to the disk and molecular cloud core, no icy planetesimal forms outside the snowline via direct collisional growth owing to the radial drift of aggregates at $\Omega t_{\rm s} \sim 1$. 
\item Dust aggregates cannot have many voids until they become large in size such that collisional compression works effectively because the small dust particles from the molecular cloud core contribute to the growth of aggregates in an earlier phase of their growth.  
\item The reason why icy aggregates suffer radial drift without growth over $\Omega t_{\rm s} \sim 1$ is that the icy region is restricted outward from the disk, where the growth rate of dust is low. 
Our analytical estimates also show that it is difficult to form an icy planetesimal in a disk having a high accretion rate $\sim 10^{-6}  \ M_{\odot} \ {\rm yr^{-1}}$, even if the internal density of aggregates is sufficiently low as $\sim 10^{-4} \ {\rm g \ cm^{-3}}$. 
\item Our results suggest that some processes that prevent small dust from growing to a macroscopic size during the disk formation stage or the streaming instability just outside the snowline may be needed to form icy planetesimals.  
\end{enumerate}

\acknowledgments
We are grateful to S. Okuzumi and S. Arakawa for their fruitful discussion and help.
This work was supported by JSPS KAKENHI (15K05266).

\bibliographystyle{aasjournal}

\bibliography{ref}




\listofchanges

\end{document}